\documentclass[useAMS,usenatbib]{mn2e}
\usepackage{mathptmx}
\usepackage{txfonts}
\usepackage[T1]{fontenc}

\usepackage{graphicx}

\newcommand\pp{{\phantom{$+$}}}
\newcommand\msol{{\cal M_{\odot}}}
\newcommand\teff{{T_{\rm eff}}}
\newcommand\lta{\mathrel{\hbox{\raise 0.6 ex \hbox{$<$}\kern
                   -1.8 ex\lower .5 ex\hbox{$\sim$}}}}
\newcommand\gta{\mathrel{\hbox{\raise 0.6 ex \hbox{$>$}\kern
                   -1.7 ex\lower .5 ex\hbox{$\sim$}}}}

\title[Isochrones for Several Mixtures of the Metals]{Models for Metal-Poor
Stars with Different Initial Abundances of C, N, O, Mg, and Si.~III.~Grids of
Isochrones for $-2.5 \le$ [Fe/H] $\le -0.5$ and Helium Abundances $Y = 0.25$
and $0.29$ at Each Metallicity}

\author[D. A. VandenBerg et al.]{
Don A.~VandenBerg \thanks{E-mail: vandenbe@uvic.ca } 
\\
Department of Physics and Astronomy, University of Victoria, 
P.O. Box 1700, STN CSC, Victoria, BC, Canada V8W 2Y2 }

\date{Accepted XXX. Received YYY; in original form ZZZ}
\pubyear{2022}

\begin{document}
\label{firstpage}
\pagerange{\pageref{firstpage}--\pageref{lastpage}}
\maketitle

\begin{abstract}

Stellar evolutionary tracks for $0.12 \le {\cal M}/\msol \le 1.0$ have been
computed for each of several variations in the abundances of C, N, and O,
assuming mass-fraction helium abundances $Y = 0.25$ and 0.29, and 11
metallicities in the range $-2.5 \le$ [Fe/H] $\le -0.5$, in 0.2 dex increments.
Such computations are provided for mixtures with [O/Fe] between $+0.4$ and
$+0.8$, for different C:N:O ratios at a fixed value of [CNO/Fe], and for
enhanced C.  Computer codes are provided to interpolate within these grids to
produce isochrones for ages $\gta 7$ Gyr and to generate magnitudes and colours
for many broad-band filters using bolometric corrections based on MARCS model
atmospheres and synthetic spectra.  The models are compared with (i) similar
computations produced by other workers, (ii) observed UV, optical, and IR
colour-magnitude diagrams (CMDs), (iii) the effective temperatures, $(V-I_C)_0$,
and $(V-K_S)_0$ colours of Pop.~II stars in the solar neighborhood, and (iv)
empirical data for the absolute magnitude of the tip of the red-giant branch
(TRGB).  The isochrones are especially successful in reproducing the observed
morphologies of optical CMDs and in satisfying the TRGB constraints.  They also
fare quite well in explaining the IR colours of low mass stars in globular
clusters, indicating that they have [O/Fe] $\approx +0.6$, though some
challenges remain.

\end{abstract}


\section{Introduction}
\label{sec:intro}

The {\it Hubble Space Telescope (HST)} UV Legacy Survey (\citealt{pmb15},
\citealt{nlp18}) obtained photometric data for 56 globular clusters (GCs) that
make it possible to distinguish between the stellar populations within them
that have different abundances of C, N, and O.  This survey employed the $F275W$
and $F336W$ filters to measure the fluxes in passbands that include spectral
features due to OH and NH, respectively, as well as the $F438W$ filter, which
samples a region of the electromagnetic spectrum with prominent CH and CN bands
(see, e.g., \citealt{mpb12}, their Fig.~11).  Indeed, the colour-magnitude
diagrams (CMDs) presented by Piotto et al.~provide a compelling demonstration
of the power of UV photometry to separate GC stars into distinct sequences, due
primarily to star-to-star variations in C:N:O at roughly constant C$+$N$+$O,
that can often be traced from the lower main sequence (LMS) to the upper
red-giant branch (RGB). Included in the Nardiello et al.~catalogues are $F606W$
and $F814W$ magnitudes, as derived from a new reduction and calibration of the
original {\it HST} Advanced Camera for Surveys (ACS) observations reported by
\citet{sbc07}.  These data provide a valuable complement to the UV photometry
insofar as the $(m_{\rm F606W}-m_{\rm F814W})_0$\ colour index places strong
constraints on the effective temperatures ($\teff$s) of stars and thereby on GC
ages (e.g., \citealt{map09}, \citealt{vbl13}, hereafter VBLC13) and star-to-star
helium abundance variations within these systems (e.g., \citealt{kbs12},
\citealt{mmp12}).

At a given metallicity, variations in the total C$+$N$+$O abundance primarily
affect the turnoff (TO) $\teff$\ and luminosity of an evolutionary track for a
fixed mass on the H-R diagram (as well the isochrones derived from them at a
fixed age), without altering the location of the RGB (see, e.g., \citealt{rc85},
\citealt{csp08}, \citealt{vbd12}).  As a consequence, the difference in the
colours of CN-weak and CN-strong giants on UV-optical CMDs is entirely a
bolometric corrections (BCs) effect; i.e., it is the abundances of C, N, and O
that were assumed in the synthetic spectra and computed BCs that matter rather
than those adopted in the stellar models.  On the other hand, TO luminosity
versus age relations depend on the CNO abundances in the interiors of stars.  As
fully appreciated by, e.g., \citet{pcs09}, \citet{ssw11}, and \citet{cmp13}, it
is therefore necessary to derive BCs from fully consistent atmospheres--interior
models and synthetic spectra in order to make meaningful comparisons between
theory and observations.  

Because such self-consistent studies were previously quite limited in scope,
\citet[hereafter Paper I]{vec22} decided to generate relatively large grids of
MARCS model atmospheres, synthetic spectra, and stellar evolutionary models for
exactly the same chemical abundances, and to explore the implications of varying
[CNO/Fe], as well as the ratio C:N:O at fixed values of [CNO/Fe].  This
exploratory study, which also examined the effects of enhancing [Mg/Fe] and
[Si/Fe] by 0.2 dex, considered three metallicties ([Fe/H] $= -2.5, -1.5, -0.5$)
and two helium abundances ($Y = 0.25, 0.29$) at each metallicity.  Moreover, BCs
for most of the broad-band filters currently in use, including those employed in
the {\it HST} UV Legacy Survey (except $F275W$) were calculated from the
synthetic spectra.  To make the best possible predictions of the fundamental
properties of LMS stars, the outer boundary conditions of the stellar interior
structures for masses in the range $0.45 \lta {\cal M}/\msol \lta 0.12$ were
derived from the MARCS model atmospheres.  Importantly, Paper I showed that both
the $\teff$s and IR colours of low-mass dwarf stars are quite dependent on the
abundances of C, N, and O.

Encouragingly, isochrones that allow for CNO abundance variations are generally
able to reproduce the morphologies of CMDs derived from the \citet{nlp18}
$F336W$, $F438W$, $F606W$, and $F814W$ observations quite well; see
\citet[hereafter Paper II]{vce22}, who considered 6 GCs with metallicities
ranging from $\approx -2.3$ (M$\,$92) to $\approx -0.7$ 
(47 Tuc).\footnote{Isochrone fits to these data, as well as to the photometry of
NGC$\,$6496 in Paper I, erroneously assumed that the $F606W$ and $F814W$
magnitudes were measured using WFC3 filters, whereas (as already noted) they are
the same ACS observations that resulted from the \citet{sbc07} investigation,
but with improvements to the reduction and calibration.  However, this error
appears to be inconsequential, judging from the differences between the
bolometric corrections given by \citet{cv14} for these two filters in the
WFC3 and ACS photometric systems.  For instance, if isochrones for [Fe/H] $=
-1.5$, [$\alpha$/Fe] $= +0.4$, and $Y = 0.25$ are transposed from the 
theoretical plane to the $(M_{F606W}-M_{F814W},\,M_{F606W})$-diagram, the
differences between the ACS and WFC3 CMDs range from $\lta 0.001$ mag along the
main sequence to $\lta 0.005$ mag along the giant branch at a fixed $M_{F606W}$
magnitude.  Clearly such small differences have no impact on the conclusions
that were drawn from the studies of $F606W$ and $F814W$ observations that were
presented in Papers I and II.}  The fits to the TO stars suggest that, 
with the exception of predicted $F336W$ magnitudes,
which appear to be too faint by $\approx 0.03$--0.04 mag, the errors in the
model fluxes at longer wavelengths are comparable with photometric zero-point
uncertainties.  Although the bluest stars along the lower RGB were universally
found to have significantly bluer $(M_{F336W}-M_{F606W})_0$ colours than any of
the model predictions, the colours spanned by most of redder giants on
UV-optical CMDs could be explained by the expected abundance differences between
CN-weak and CN-strong stars.  It would clearly be worthwhile to carry out
similar investigations of other GCs given the great diversity of their CMDs, as
presented by \citet{pmb15}.  Such work may reveal other significant challenges
to our understanding of these objects, in addition to constraining the absolute
chemical abundances which characterize the multiple stellar populations that
they contain.  

For the CMD analyses presented in Paper II, several grids of evolutionary
sequences and isochrones were computed specifically for the current best
estimates of the metallicities of the clusters that were studied.  A more
efficient way of accomplishing the same thing is to generate grids of tracks
for a fine spacing in [Fe/H] and $Y$ for each of the most relevant mixtures of
the light elements that were assumed in Paper I and to provide the means to 
interpolate in those grids so as to obtain isochrones for the various mixtures
of the metals, [Fe/H] values, He abundances, and ages of interest.  It is the
primary purpose of the present study to provide such grids of stellar models
for general use.  

The next section briefly describes the calculation of the evolutionary tracks
and, for a metallicity near the middle of the range spanned by GCs (specifically
[Fe/H] $= -1.3$), compares Victoria-Regina (hereafter V-R) isochrones
with those published by other workers.\footnote{Whereas the tracks were
generated using a code developed over the years by D.A.V.~and can
be considered ``Victoria" stellar models, isochrones derived from them should be
referred to as ``Victoria-Regina" isochrones to acknowledge the use of the
sophisticated interpolation codes that were devloped by P.~A.~Bergbusch (see
\citealt{vbd12}, and references therein), who worked at the University of Regina
prior to his retirement.}  Section 3 presents and discusses the application of
the present isochrones to recent IR CMDs of GCs, in addition to showing that (i)
the predicted $\teff$ and colour scales are consistent with those of Pop.~II
dwarfs in the solar neighbourhood with accurate distances and (ii) the models
satisfy current constraints on the relations between the absolute $I_C$
magnitude of the RGB tip with $(V-I)_C$ colour, or with [Fe/H], very well.  A
brief summary of the main results of this investigation is provided in Section 4.

\section{The Computation of Stellar Evolutionary Sequences}
\label{sec:models}

Grids of evolutionary tracks for [Fe/H] $= -2.5$ to $-0.5$, in steps of 0.2 dex,
were computed for the mixtures of C, N, and O that are listed in
Table~\ref{tab:t1}.  In the case of the {\tt a4s21} mix, the $\log N_i$
abundances, on the scale $\log N_{\rm H} = 12.0$, correspond to the adopted
\citet{ags09} solar abundances with the addition of a 0.4 dex enhancement of O
(and all of the other so-called ``alpha elements", though they are not tabulated
explicitly) to be consistent with [$\alpha$/Fe] $= +0.4$.  The others assume the
same abundances except for the adjustments, in dex, that are specified by upward
or downward pointing arrows.  For instance, the {\tt a4CNN} mixture has a lower
abundance of carbon by 0.3 dex but a higher nitrogen abundance by 1.13 dex,
resulting in $\log N_{\rm C} = 8.13$ and $\log N_{\rm N} = 8.96$, respectively.
The tabulated numbers are equivalent to saying that the {\tt a4CNN} mix assumes
[C/Fe] $= -0.3$ and [N/Fe] $= +1.13$.  Clearly, the CNO abundances at any
metallicity can be obtained simply by adding the [Fe/H] value of interest to
the adjusted $\log N_i$ values.  

\begin{table}
\centering
\caption{The Adopted Abundances of C, N, and O} 
 \label{tab:t1}
\smallskip
\begin{tabular}{lccccc}
\hline
\hline
\noalign{\smallskip}
 Mixture & & & & & \\
 Name$(^a)$  & He & C & N & O & [CNO/Fe] \\
\noalign{\smallskip}
\hline
\noalign{\smallskip}
 a4s21$^{b}$  & 11.00 & \pp8.43 & \pp7.83 & \pp9.09 & $+0.28$ \\
 a4CNN  & \bf{--} & $\downarrow\,0.3$ & $\uparrow\,1.13$ & \bf{--} & $+0.44$ \\
 a4ONN   & \bf{--} & $\downarrow\,0.8$ & $\uparrow\,1.48$ & $\downarrow\,0.8$ &
   $+0.44$ \\
 a4xCO   & \bf{--} & $\uparrow\,0.7$ & \bf{--} & $\uparrow\,0.2$ & $+0.61$ \\
 a4xO$\_$p2  & \bf{--} & \bf{--} & \bf{--} & $\uparrow\,0.2$ & $+0.44$ \\ 
 a4xO$\_$p4  & \bf{--} & \bf{--} & \bf{--} & $\uparrow\,0.4$ & $+0.62$ \\ 
\noalign{\smallskip}
\hline
\noalign{\smallskip}
\end{tabular}
\begin{minipage}{1\columnwidth}
$^{a}$~These mixtures are among those considered in Papers I and II. \\
$^{b}$~In this study, the name {\tt a4xO\_p0} is sometimes used instead of
{\tt a4s21} to represent \citet{ags09} abundances that have been adjusted
to include a 0.4 dex enhancement of oxygen and the other$\alpha$ elements.
This is a more meaningful name when comparing stellar models for this mixture
with those that assume higher O abundances (i.e., {\tt a4xO\_p2} and
{\tt a4xO\_p4}). \\
\phantom{~~~~~~~~~~~~~~~}
\end{minipage}
\end{table}

Paper I considered more than a dozen different mixtures of the metals, but most
of the mixes for [CNO/Fe] $= +0.28$, with various ratios of C:N:O, were
dropped from consideration because the eclipsing binaries in GCs favour stellar
models that assume [O/Fe] $\gta +0.6$ (see Paper II), in good agreement with
spectroscopic determinations of the O abundances in solar neighbourhood Pop.~II
stars (e.g., \citealt{fna09}, \citealt{ncc14}, \citealt{ans19}).  In fact, if
[CNO/Fe] were as low as $+0.28$, assuming the \citet{ags09} scale of solar
abundances, the maximum possible value of [N/Fe] would be $\approx +1.30$ if
all of the carbon and oxygen implied by an initial mixture with [C/Fe] $= 0.0$,
[N/Fe] $= 0.0$ and [O/Fe] $=$ [$\alpha$/Fe] $= +0.4$ were converted to nitrogen.
Since GCs are typically observed to have C$+$N$+$O $\approx constant$ (e.g.,
\citealt{ssb96}, \citealt{cm05}) and for some fraction of the member stars to
have [N/Fe] $\gta +1.5$ (e.g., \citealt{bcs04}, \citealt{sbh05}, \citealt{cbs05}),
they apparently have [CNO/Fe] $\gta 0.44$.  This value is obtained, for
instance, if the initial mixture has [C/Fe] $=$ [N/Fe] $= 0.0$ and [O/Fe]
$= +0.6$ (see Table~\ref{tab:t1}).  Note that, as discussed in Paper I, a gas
with initial C, N, and O abundances similar to those of the {\tt a4xO\_p2} mix
can be expected to have close to the C:N:O ratios given by the {\tt a4CNN} or
{\tt a4ONN} cases, in turn, if it were subjected to efficient CN-cycling or to
ON-cycling at high temperatures.  The others make it possible to study the
photometric consequences of a mix with high carbon ({\tt a4xCO}) or enhanced
oxygen ({\tt a4xO\_p4}).

All of the stellar models that were generated for the present series of papers
were computed using the code described by \citet[and references
therein]{vbd12}.  It is fully up-to-date insofar as the basic physics
incorporated in it is concerned, except that the diffusion of the metals has
not been considered.  As reported by VandenBerg et al., the gravitational
settling of helium, which is of considerable importance for the morphologies
of evolutionary tracks prior to the turnoff and for predicted TO luminosity
versus age relations (also see \citealt{vrm02}) has been treated, along with
extra mixing below envelope convection zones when they occur.\footnote{In the 
Victoria code, free parameters associated with the extra mixing formalism (see
\citealt{vbd12}) were set by following the nucleosyntheis of the LiBeB elements
and requiring that a Standard Solar Model reproduce the observed surface Li
abundance of the Sun at the solar age.}  Fully diffusive computations by e.g.,
\citet{rmr02} have shown that additional mixing at the inner boundaries of
surfrace convection zones, possibly due to turbulence, appears to be necessary
to satisfy such empirical constraints as the observed Li abundances in field
halo dwarfs (\citealt{sms84}, \citealt{rnb99}) and the measured abundance
variations between TO and lower RGB stars (see, e.g., \citealt{gbb01},
\citealt{rc02}).  Indeed, more recent studies by \citet{nkr12}, and
\citet{gnk14}, among others, have found that TO stars in GCs, which should show
the strongest signature of diffusion, have surface iron abundances that are only
$\sim 0.1$ dex lower than those derived for stars just beginning their ascent
of the RGB.  This is far smaller than the 0.5--0.7 dex variation that is
predicted by fully diffusive models that neglect extra mixing at very low
metallicities (\citealt{rmr02}).

Although radiative accelerations probably do not play a very significant role in
our understanding of the evolution of GC stars (see the discussion of this point
by \citealt{phc21}), the gravitational settling of the metals should, in
principle, be taken into account.  However, in practice, the neglect of this
process will have no more than minor consequences, at least for stellar models
that are applied to GC CMDs.\footnote{The diffusion of the metals should not be
ignored in more exacting investigations, such as comparisons of predicted
sound speed profiles for the solar interior with that inferred from helioseismic
observations (e.g., \citealt{bsp05}), studies of metal abundance variations
between MS and lower RGB stars in GCs (e.g., \citealt{kgr07}), or the
determination of the minimum mass, at a fixed metallicity, that is able to
retain a convective core throughout its core H-burning phase (e.g.,
\citealt{mrr04}).}  As a general rule of thumb, a 0.3 dex (i.e., a factor of
two, or a 100\%) increase of the C$+$N$+$O abundance in the nuclear burning
regions of low mass, metal-poor stars results in a reduction of the predicted
age at a given TO luminosity by $\sim 1$ Gyr.  Since the settling of the metals
will increase the central abundances of CNO and the other metals by a few to
several percent over MS lifetimes of $\gta 10$ Gyr, it can be expected that this
process alone will reduce TO ages by $\sim 0.1$ Gyr (at most).  This is
insignificant compared the uncertainties associated with cluster distances and
chemical abundances.  The settling of helium is much more important for GC ages
because He is so abundant compared with any of the metals. 

\begin{figure}
\begin{center}
\includegraphics[width=\columnwidth]{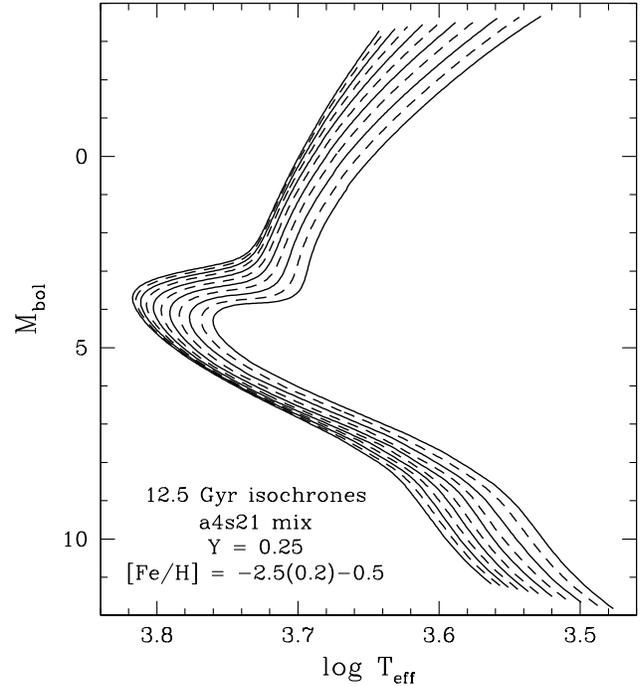}
\caption{A plot on the H-R diagram of 12.5 Gyr isochrones for $Y=0.25$ and
[Fe/H] values from $-2.5$ to $-0.5$ in steps of 0.2 dex.  These particular
isochrones are for the {\tt a4s21} mix.}
\label{fig:f1}
\end{center}
\end{figure}

\begin{figure*}
\begin{center}
\includegraphics[width=0.95\textwidth]{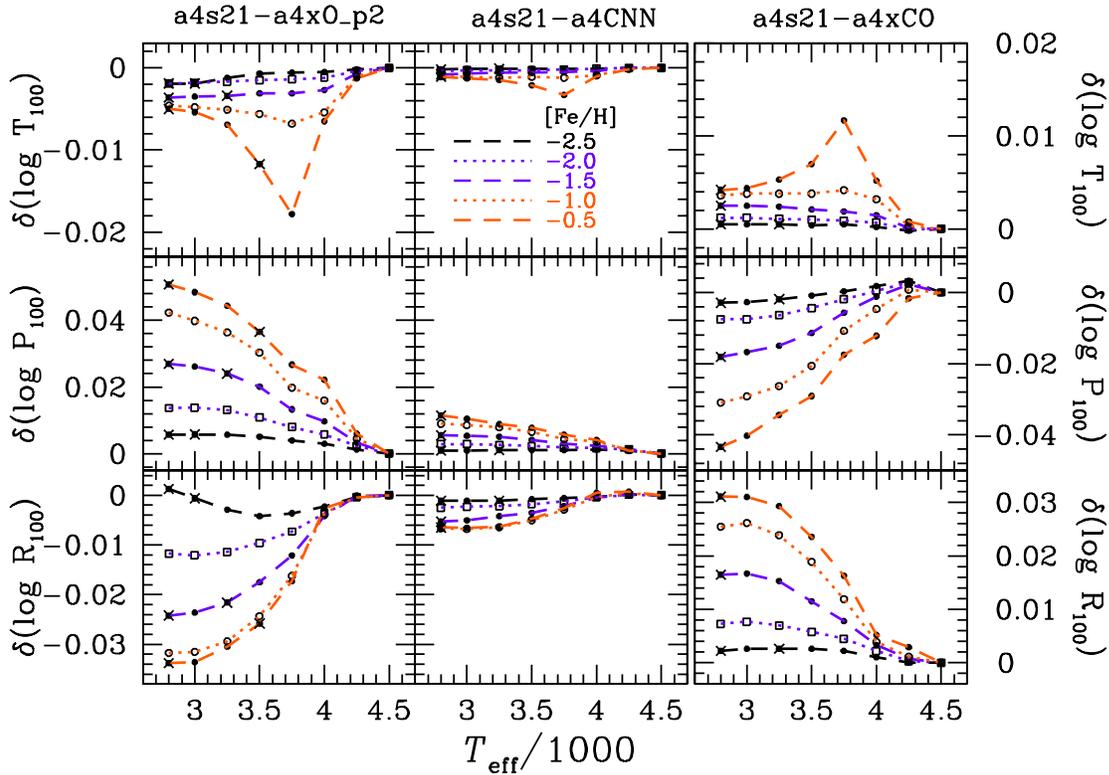}
\caption{Differences of the temperatures, pressures, and radii at $\tau = 100$
(top, middle, and bottom rows of plots, respectively) between the predictions
from MARCS model atmospheres for the {\tt a4s21} mix and, in turn, the
{\tt a4xO\_p2}, {\tt a4CNN}, and {\tt a4xCO} mixtures (left-hand, middle, and 
right-hand columns of plots) as a function of $\teff$ and the indicated [Fe/H]
values.  All model atmospheres were computed for $\log\,g = 5.0$.  Crosses
indicate those atmospheres that failed to converge; these points were obtained
by interpolating in, or extrapolating, the results from the converged models.
Note that the ordinate values along the right-hand edge apply only to the
right-hand column of plots.}
\label{fig:f2}
\end{center}
\end{figure*}

The veracity of these remarks is borne out by the comparison given below of V-R
isochrones with those from the latest BaSTI grids (\citealt{phc21}).  Even 
though the latter, but not the former, follow the settling of the metals, both
sets of isochrones predict virtually identical TO luminosities when the same
[Fe/H], [CNO/Fe], and age are assumed.  It should be appreciated, however, that
there are differences in the respective treatments of diffusive processes that
may have compensated for the expected small offset of these computations.
Whereas a code very similar to the one written by \citet{tbl94} is used in the
the BaSTI stellar evolution program to solve the applicable transport equations
(\citealt{bur69}), somewhat improved versions of the subroutines developed by
\citet[see the Appendix of their paper]{pm91} are employed in the Victoria code.

There are some differences in the predicted $\teff$s of TO stars, but they are
relatively minor ($\sim 60$~K) and they may be due in part to other differences
between the Victoria and BaSTI evolutionary programs besides the treatments of
diffusion.  In fact, because the predicted $\teff$ scale is subject to so many
uncertainties that are hard to evaluate, such as known deficiences in the
treatments of convection and the atmospheric boundary condition, one must rely
on comparisons with empirical constraints to evaluate the reliability of the
temperatures predicted by any stellar models.  As also shown below, V-R 
isochrones provide rather good fits, not only to the temperatures of field
Pop.~II stars as derived from the Infrared Flux Method (\citealt{crm10}), but
also to the TO colours and morphologies of observed CMDS, especially when
derived from the measured fluxes in optical passbands.

As in all previous presentations of V-R isochrones since the study by
\citet{vbf14}, MARCS model atmospheres at an optical depth $\tau = 100$ have
been attached to the interior structures of stellar models for masses $\lta 0.45
\msol$ in order to make the best possible predictions of the $\teff$s of LMS
stars.  In higher mass models, the photosphere was taken to be the outer
boundary and the pressure at $T=\teff$ was determined by integrating the
hydrostatic equation in conjunction with an assumed $T$--$\tau$ relation ---
specifically, the fit given by \citet{vp89} to the semi-empirical solar
atmosphere derived by \citet{hm74} --- from very small optical depths to the
photospheric value of $\tau$.  The justification for this procedure, the
transition between the two mass regimes, and the whole issue of the atmospheric
boundary condition has already been discussed quite extensively by
\citet{vbf14}; consequently, little else needs to be mentioned here concerning
this aspect of the models.  As illustrated in Figure~\ref{fig:f1}, isochrones
generated from the evolutionary tracks that have been computed for this project
(for the {\tt a4s21} mix, in this particular case) are very smooth over the full
range in luminosity between the LMS and the RGB tip.  Those for the other metal
abundance mixtures listed in Table~\ref{tab:t1} look qualitatively very similar,
though there are small systematic differences between them as a function of
$\teff$ and luminosity.

The only modeling results that have not been discussed previously involve the
dependence of the atmospheric properties at $\tau = 100$ on the assumed mix of
the metals and their consequences for predicted $\teff$s.  Figure~\ref{fig:f2}
plots the differences in $\log T$, $\log P$ and $\log R$ at $\tau = 100$ between
the {\tt a4s21} mix and the {\tt a4xO\_p2}, {\tt a4CNN}, and {\tt a4xCO}
mixtures in the left-hand, middle, and right-hand panels, respectively, as a
function of $\teff$ and [Fe/H].  (B.~Edvardsson kindly computed additional sets
of model atmospheres for [Fe/H] $= -2.0$ and $-1.0$ and $\log\,g \ge 4.7$ for
each of the metal abundance mixtures in Table~\ref{tab:t1} to complement those
for [Fe/H] $= -2.5$, $-1.5$, and $-0.5$, which were produced for Paper I.)  Not
surprisingly, the differences increase with increasing [Fe/H] and usually with
decreasing $\teff$, though maxima or minima occur at $\sim 3800$~K in the loci
for $\delta\log T_{\rm 100}$ at the highest metallicities (likely associated
with the formation of TiO and other molecules involving O and C).  Oxygen
enhancements result in higher values of $T_{\rm 100}$ and $R_{\rm 100}$, but
lower values of $P_{\rm 100}$, as shown in the left-hand panels, while the
reverse occurs if the abundances of C and O are increased by large amounts, as
in the {\tt a4xCO} mix (see the right-hand panels).  In contrast with these two
cases, the structural properties change only slightly if a mixture has higher
N and lower C similar to the abundances found in CN-strong stars (the
{\tt a4CNN} mix) as compared with model atmospheres for the {\tt a4s21} mixture, 
which is relevant to CN-weak stars; see the results in the middle column of
panels.  (Note that there are differences in the [CNO/Fe] values of the various
mixtures considered in this figure; see Table~\ref{tab:t1}.)

\begin{figure}
\begin{center}
\includegraphics[width=\columnwidth]{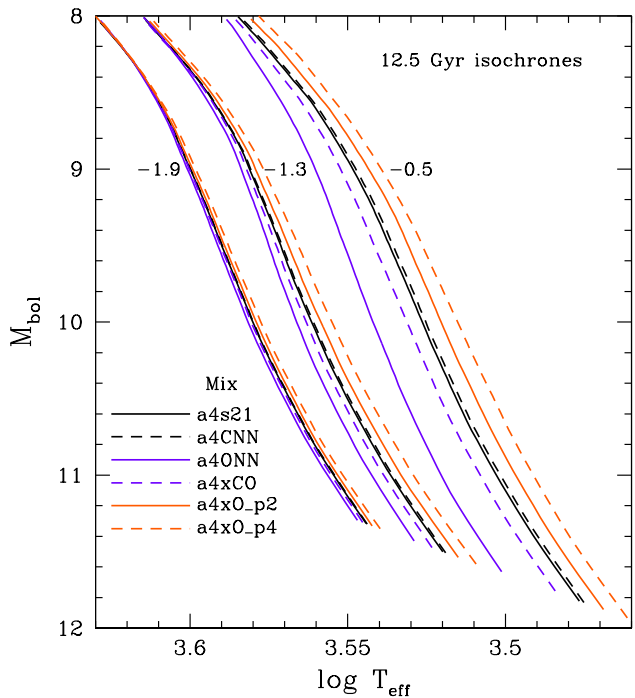}
\caption{Comparison on the H-R diagram of the LMS portions of 12.5 Gyr
isochrones for the indicated mixtures of the metals, assuming $Y = 0.25$ and
[Fe/H] $= -1.9, -1.3$, and $-0.9$ (the three groups of loci in the direction
from left to right).}
\label{fig:f3}
\end{center}
\end{figure} 

Figure~\ref{fig:f3} compares the LMS portions of some of the isochrones that
have been computed for this project.  It is clear that reduced or increased
abundances of oxygen have significant effects on the $\teff$s of LMS stars,
as indicated by the loci for the {\tt a4xO\_o4} and the {\tt a4ONN} mixtures.
The temperature offsets are particularly large at high metallicities, though
they are still substantial down to [Fe/H] $\lta -1.6$.  Given the results in
the previous figure, the LMS models will have higher $\teff$s when the
atmospheres have lower pressures and larger radii at large optical depths, and
vice versa.  Worth pointing out are the offsets between the solid curves in
orange, for the {\tt a4xO\_p2} mix, and the dashed curves in purple, for the
{\tt a4xCO} mixture.  As the only difference between these cases is a 0.7 dex
enhancement of the carbon abundance, which is not large enough for C to be
more abundant than O, Fig.~\ref{fig:f3} shows that carbon can shift LMS loci
to appreciably higher temperatures, at a fixed luminosity, if it has a high
abundance.

\subsection{Comparisons with BaSTI and DSEP Isochrones}
\label{subsec:compare} 

\citet{phc21} have recently released the latest edition of BaSTI stellar models
for [$\alpha$/Fe] $= +0.4$,\footnote{http://basti-iac.oa-teramo.inaf.it} which
involve a number of improvements or modifications relative to those published
previously (e.g., \citealt{pcs13}, and references therein), including the
settling of He and the metals (as already noted), an updated reference solar
distribution of the metals (from \citealt{cls11}), and new treatments of the
atmospheric boundary condition and bolometric corrections (also see
\citealt{hpc18}).  Because those computations assume a different reference mix
than the Victoria models, interpolations have been made in the grids of
evolutionary sequences for the {\tt a4s21} and {\tt a4xO\_p2} mixtures in order
to obtain a set of tracks for the same C$+$N$+$O abundance as the BaSTI models.
This constraint can be satisfied if the value of $\log N_{\rm O}$ given by the
\citet{ags09} scale of chemical abundances is increased by 0.08 dex (in
addition to the 0.4 dex enhancement implied by [$\alpha$/Fe] $= +0.4$, before
the abundances are adjusted to the desired [Fe/H] value).  As noted in
Table~\ref{tab:t1}, the {\tt a4s21} and {\tt a4xO\_p2} mixes have [O/Fe]
$= +0.4$ and $+0.6$, respectively, and it is a straightforward exercise to
interpolate in them to obtain evolutionary tracks for [O/Fe] $= +0.48$.
Although there are differences in the ratio C:N:O, it is the total C$+$N$+$O
abundance that is of primary importance for TO luminosities and temperatures.

\begin{figure}
\begin{center}
\includegraphics[width=\columnwidth]{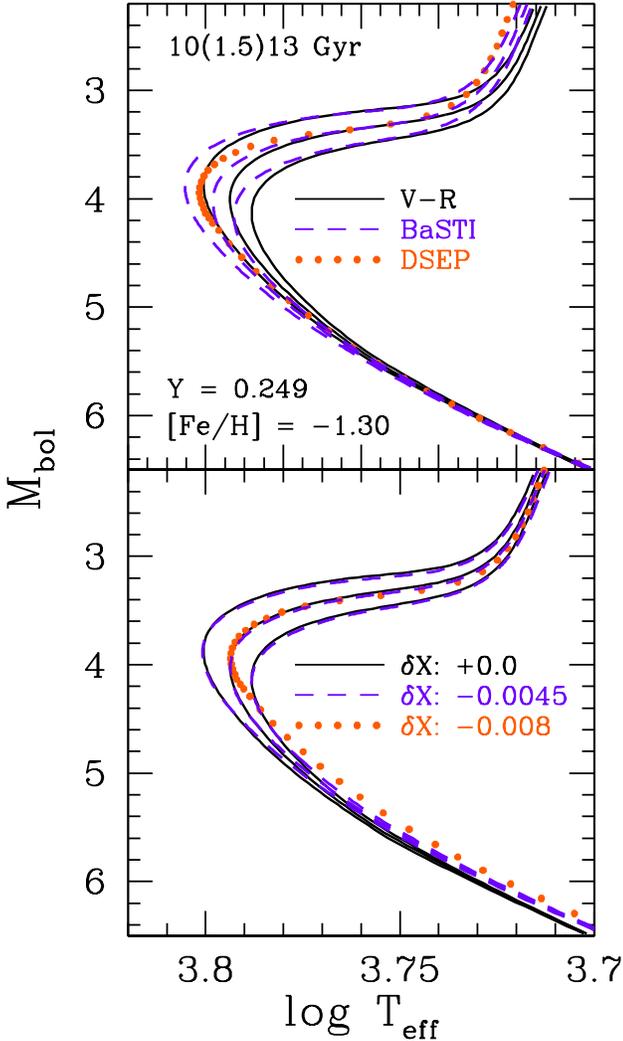}
\caption{{\it Top panel:} Comparison of V-R and BaSTI isochrones for the
the indicated metallicity, helium abundance, and ages.  Also shown is an 11.5
Gyr isochrone for the same [Fe/H] value, but for $Y = 0.248$.  {\it Bottom
panel:} As in the top panel, except that the isochrones have been adjusted 
horizontally by the indicated $\delta\,X$ values, where $X = \log\teff$.}  
\label{fig:f4}
\end{center}
\end{figure} 

The top panel of Figure~\ref{fig:f4} compares V-R and BaSTI isochrones
for $Y = 0.249$, [Fe/H] $= -1.30$, $\log(N_{\rm C}+N_{\rm N}+N_{\rm O}) = 7.96$,
and ages of 10.0, 11.5, and 13.0 Gyr.  Although they overlay one another along
the MS (at $M_{\rm bol} \gta 5.8$), the BaSTI isochrones have somewhat hotter
TOs and RGBs than the V-R isochrones.  However, the more interesting plot is
the one in the bottom panel.  If the BaSTI isochrones are adjusted to cooler
temperatures by only $\delta\log\teff = 0.0045$, their TOs superimpose those
of the V-R counterparts almost exactly, and there are only slight differences
between the two along the subgiant branch (SGB) and lower RGB.  Of course, the
MS portions of the two sets of isochrones no longer coincide if such a
horizontal adjustment is applied to the BaSTI loci.  However, the main point of
this plot is simply to show that the V-R and BaSTI isochrones predict exactly
the same turnoff luminosities at the same ages when both sets assume almost the
same abundances.  There are, after all, small differences in the abundances of
most of the metals, including, e.g., Ne, Mg, and Si, in the solar distributions
of the elements tabulated by \citet{ags09} and \citet{cls11}. 

Also plotted in Fig.~\ref{fig:f4} is a DSEP isochrone
({\citealt{dcj08})\footnote{http://stellar.dartmouth.edu/models} for [Fe/H]
$= -1.30$, $Y = 0.248$, and an age of 11.5 Gyr.  (As Dotter et al.~adopt a
slightly different enrichent law, $\Delta\,Y/\Delta\,Z$, for the variation of
$Y$ with the mass-fraction abundance of the metals, $Z$, the helium abundance of
the DSEP isochrone at the value of $Z$ corresponding to [Fe/H] $= -1.30$ is
smaller by $\delta\,Y = 0.001$ that that of the BaSTI isochrone.)  Once again,
it is the bottom panel that is the most instructive one.  It reveals that the
morphology of the DSEP isochrone in the vicinity of the turnoff is quite
different from those of the others that have been plotted; in particular, the
isochrone has an odd shape in the luminosity range $4.6 \gta M_{\rm bol} \gta
4.0$.  Furthermore, because the DSEP models adopted the reference solar
abundances given by \citet{gs98}, which has higher C, N, and O than the
\citet{cls11} scale, one would have expected that at, the same TO luminosity,
the DSEP isochrone would predict a younger age, by $\approx 0.2$ Gyr, than a
BaSTI isochrone for the same chemical abundances.  It should also be somewhat
cooler at the TO, but the top panel shows that this is not the case.  The
apparent inconsistencies cannot be attributed to differing treatments of
diffusion because the methods described by \citet{tbl94} have been employed in
both the BaSTI and DSEP codes to follow the settling of helium and the metals.
However, there must be some differences in the physics incorporated in these
evolutionary codes in order to explain the offsets in both the predicted
luminosities and temperatures of the respective isochrones.

\begin{figure*}
\begin{center}
\includegraphics[width=\textwidth]{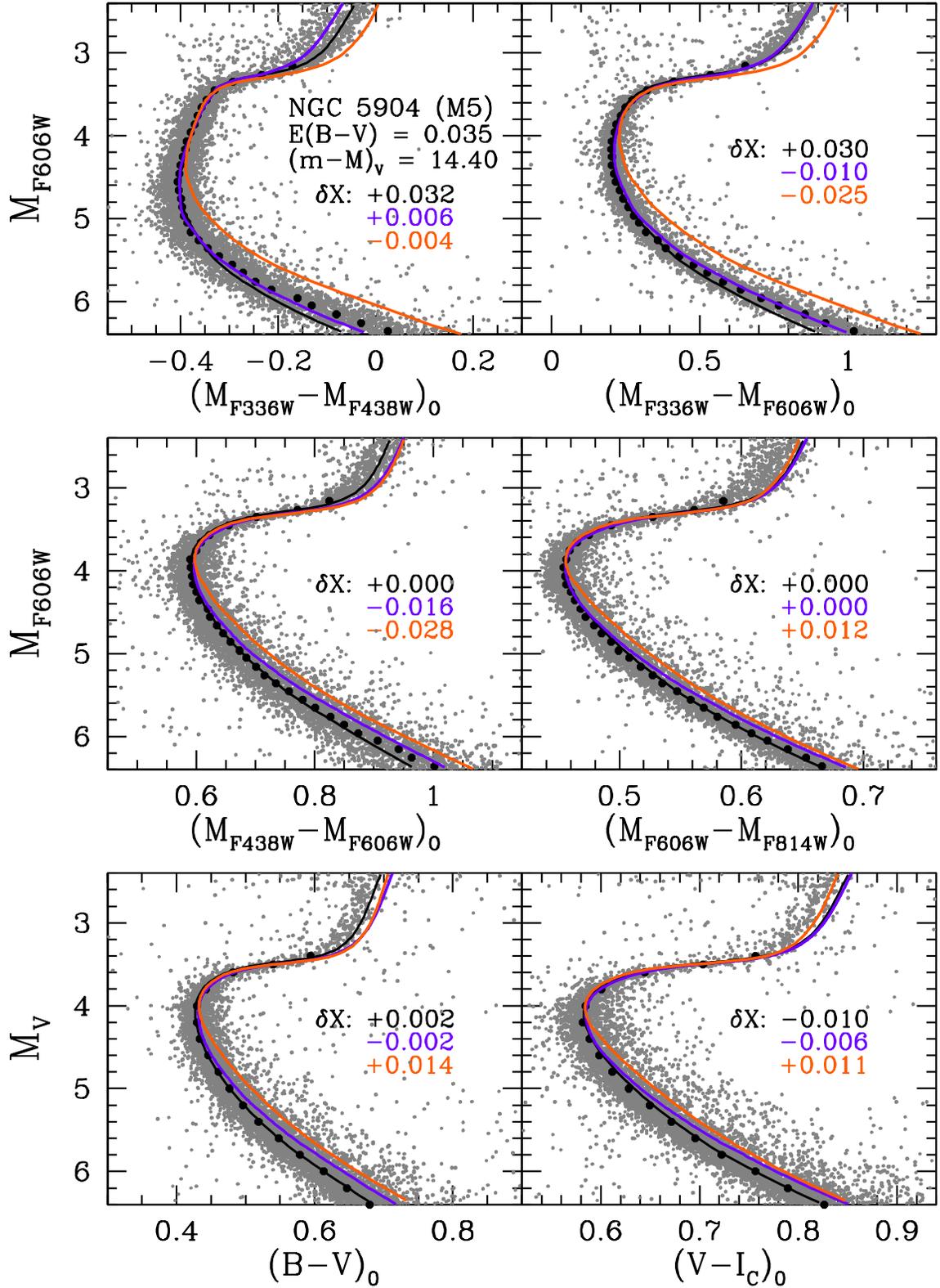}
\caption{Comparisons of the same isochrones that appear in the previous figure
with the CMDs of M$\,$5 that have been derived from {\it HST} UV Legacy
Survey observations (the top four panels) and from ground-based $BVI_C$
photometry (the bottom two panels); these data were released for general use
by \citet{nlp18} and \citet{spz19}, respectively.  The adopted cluster 
reddening and the $V$-band apparent distance modulus are specified in the
top left-hand panel.  The $\delta\,X$ values in each panel indicate the colour
offsets that were applied to the isochrones in order to obtain best fits to the
turnoff stars.  The filled circles in black represent median fiducial points 
for the different CMDs, while the solid curves in black, purple, and orange
represent, in turn, the V-R, BaSTI, and DSEP isochrones.}
\label{fig:f5}
\end{center}
\end{figure*} 

There is ample evidence that the V-R isochrones provide good fits to the 
CMDs of GCs (e.g., \citealt{vbf14}, \citealt{vd18}, Paper II), while DSEP
isochrones have some difficulties in this regard (see, e.g., \citealt{dsa10},
their Figs.~4 and 5).  Indeed, the latest BaSTI isochrones (\citealt{phc21})
also do a better job of reproducing cluster CMDs than DSEP models, though they
appear to be somewhat less successful than V-R computations in explaining
$BVI_C$ observations, or {\it HST} photometry of GCs at optical wavelengths.
This is shown in Figure~\ref{fig:f5}, where the same 11.5 Gyr isochrones that
were intercompared in Fig.~\ref{fig:f4} are fitted to various CMDs for
NGC$\,$5904 (M$\,$5), which has a metallicity close to [Fe/H] $= -1.30$ (e.g.,
\citealt{cbg09}).  If $E(B-V) = 0.035$ is adopted for M$\,$5, which corresponds
to the average line-of-sight reddening from the studies by \citet{sfd98} and
\citet{sf11}, the cluster TO stars can be matched quite well by 11.5 Gyr
isochrones if $(m-M)_V \approx 14.40$, which is close to the apparent
distance modulus implied by ZAHB models (see, e.g., VBLC13).\footnote{The main
purpose of Fig.~\ref{fig:f5} is simply to illustrate how the three isochrones
compare with each other and with various CMDs for M$\,$5.  These intercomparisons
are not contingent upon the assumed age and cluster distance.  In other words,
plots of younger or older isochrones would reveal the same morphological
differences between them and the same difficulties when compared with the M$\,$5
CMD.  For instance, had isochrones for an age of 10.5 Gyr been considered, which
would have implied a larger distance modulus by $\approx 0.10$ mag, the CMDs in
the various panels would be shifted upwards by only one half of the magnitude
differences between adjacent tick marks along the respective $y$-axes and it
would have been necessary to adjust the predicted colours slightly to the red 
in order for 10.5 Gyr isochrones to reproduce the intrinsic colours of the TO
stars.}  The sources of the $F336W$, $F438W$, $F606W$, and $F814W$ observations,
on the one hand, and the $BVI_C$ data, on the other, are \citet{nlp18} and
\citet{spz19}.

Note that, in this investigation, the $E(B-V)$ values that are mentioned in
the text or in figure legends are so-called ``nominal" reddenings; i.e., they
are applicable to early-type stars, which is the usual convention for many
reddening determinations in the literature, including those derived from the 
\citet{sfd98} dust maps.   The actual colour excess, $E(\zeta-\eta)$ for filters
$\zeta$ and $\eta$, that applies to the stars in a given GC can be calculated
to sufficient accuracy (especially when the reddening is low) using the values
of $R_\zeta$ and $R_\eta$ given by \citet[their Table A1]{cv14}, where
$R_\zeta = A_\zeta/E(B-V)$, $A_\zeta$ is the extinction and $E(B-V)$ is the
nominal reddening (similarly for $R_\eta$).  Since the reddening produced by a
given amount of dust is less for stars of later spectral types, it follows that
$R_B - R_V < 1$; to be specific, Casagrande \& VandenBerg's Table A1 gives
$R_B -R_V = 0.922$ for F-type TO stars.  One could equivalently specify the
actual $E(B-V)$, but then the colour excess $E(\zeta-\eta)$, in general, would 
have to be calculated by multiplying this value of $E(B-V)$ by
$(R_\zeta-R_\eta)/(R_B-R_V)$.  It is simpler just to give the nominal 
reddening and to adopt the $R_\lambda$ values, as tabulated, to derive the
colour excesses of interest.  

Furthermore, apparent distance moduli as measured in the $V$ magnitude are
generally provided instead of, say, $(m-M)_{F606W}$ when fitting isochrones
to the $(M_{F606W}-M_{F814W})_0,\,M_{F606W}$ CMD, because they are commonly used
to specify GC distances.  However, they can be easily converted to
$(m-M)_\zeta$, where $\zeta$ is the filter of interest, using $(m-M)_\zeta =
(m-M)_V + (R_\zeta - R_V)E(B-V)$, where $E(B-V)$ is the nominal reddening and
$R_\zeta$ has the value provided by \citet{cv14}.  Note that the standard $R_V =
3.1\,E(B-V)$ reddening law was assumed in these computations.   

To match the cluster TO stars, the various isochrones had to be shifted in 
colour by the amounts indicated by the $\delta\,$X values. The loci in black,
purple, and orange represent, in turn, the V-R, BaSTI, and DSEP models to be
consistent with the colours adopted in Fig.~\ref{fig:f4}.  (It should be kept in
mind that comparisons with observed photometry involve both the predicted
$\teff$s and the adopted bolometric corrections.)  In the case of the V-R
isochrones, the BCs were derived from the the transformations provided by
\citet{cv14}, but with the adjustments reported in Paper I to be consistent
with recent improvements to MARCS model atmospheres and synthetic spectra 
(though they are insignificant for passbands in the optical and IR regions of
the electromagnetic spectrum).  The BaSTI and DSEP isochrones employ different
BCs, as described in their papers; users are able to select the synthetic
magnitudes of interest when requesting stellar models from their respective web
sites (see footnotes 5 and 6). 

Fig.~\ref{fig:f5} shows that V-R isochrones require small redward colour offsets
to fit the TO stars in the top two panels, but very little, if any, adjustments
to accomplish similar fits in the bottom four panels.  Indeed, these models
superimpose the median fiducial sequences of M$\,$5 (the filled circles in
black) in these panels very well, though the models become systematically too
blue at faint magnitudes in CMDs that involve $F336W$ magnitudes.  These results
offer strong support for both the predicted $\teff$s and the colour 
transformations based on MARCS synthetic spectra, though there would appear to
be some deficiencies in the MARCS models in the UV that cause the apparent
zero-point and systematic differences that are apparent in the top two panels.

Interestingly, the BaSTI isochrones provide quite agreeable fits to UV-optical
colours (see the top row of panels in Fig.~\ref{fig:f5}), though they are less 
successful than the V-R models in matching CMDs that involve the magnitudes
measured in redder filters (see the bottom four panels).  In all instances, the
zero-point adjustments that were applied to the colours of the BaSTI models are
sufficiently small to be within the zero-point uncertainties associated with
both the observed and predicted photometry. The apparent systematic difficulties
are probably mostly associated with the predicted temperatures and the BCs
that are used to transpose the models from the theoretical H-R diagram to
observed CMDs than to, e.g., the adopted chemical abundances.  Indeed, if
the BaSTI isochrones were adjusted to somewhat bluer colours in order that
they superimpose the observed MS fiducials of M$\,$5 at $\gta 1.5$ mag below
the TO, the overall consistency between theory and observations would likely be
improved --- though any horizontal shifts that are applied to the entire
isochrones would necessarily result in small discrepancies between the predicted
and observed TO colours.  Finally, it is quite clear from Fig.~\ref{fig:f5}
that DSEP isochrones are not able to match the M$\,$5 CMDs very well, which can
be attributed in part to their unusual morphology in the vicinity of the TO
(recall Fig.~\ref{fig:f4}) and likely to their adopted colour transformations.

There are a couple of scientific results in Fig.~\ref{fig:f5} that are worth
pointing out.  The assumed abundances in all of the isochrones that appear in
this plot are similar to those found in CN-weak stars, which are predicted to
have blue $(M_{F336W}-M_{F438W})_0$ colours.  Paper II was unable to offer an
explanation for the bluest giants along the lower RGB in M$\,$5, or similar
populations that seem to be present in all other GCs, other than to suggest that
they have [N/Fe] $< 0.0$.  As shown in the top, left-hand panel, the BaSTI
isochrone also fails to match these giants in M$\,$5 though the observations
present less of a  problem for the BaSTI models than for the V-R isochrone.
Higher $Y$ would reduce the discrepancies, though the helium abundance would
likely have to be appreciably higher than $Y = 0.30$ to eliminate this
problem (see Paper II).  

In addition, all of the isochrones predict intrinsic $(V-I)_C$ and
$M_{F606W}-M_{F814W}$ colours that are similar to those found in only the
reddest giants (see the relevant plots in Fig.~\ref{fig:f5}).  The difficulty
here is that, as discussed in Paper II, such colours appear to be insensitive to
variations in the abundances of C, N, and O, which would seem to leave high $Y$
as the most likely explanation for bluer colours.  However, the horizontal
branch (HB) of M$\,$5 does not have an extended blue tail; consequently,
star-to-star variations in $Y$ are probably not very large.  This is suggested
by recent studies of the M$\,$3 HB (\citealt{dvk17}, \citealt{tdc19}), which is
morphologically quite similar to the HB of M$\,$5.  Although so-called
``chromosome maps" appear to favor larger He abundance variations than those
inferred from HB simulations, especially in the case of M$\,$3 ({\citealt{mmr18}),
Tailo et al.~present a number of quite compelling arguments against this
possibility.  Errors in the model $\teff$s may be partly responsible for the
discrepancies between the predicted and observed $(V-I)_C$ colours (and similar
{\it HST} colours), but the indications for such errors are less apparent when
other colours are considered.  It may be worthwhile to explore the implications
of even more extreme chemical abundance variations (in particular, [N/Fe] $>
+1.5$) than those examined in Paper II.


\section{Empirical Constraints}
\label{sec:constraints}

CMD studies, such as those just discussed, provide useful tests of stellar
models, especially if the basic cluster parameters (distance, reddening,
metal abundances) are derived by independent methods; i.e., if such 
fundamental properties are not based on the fits themselves of the isochrones
to the photometric data (as in the work by, e.g., \citealt{wsv17},
\citealt{gkm21}).  For instance, recent papers that have used V-R isochrones to
derive cluster ages (e.g., \citealt{vdc16}, {\citealt{dvk17}), generally adopted
distance moduli that were obtained by fitting zero-age horizontal branch (ZAHB)
loci to their HB populations on the assumption of reddenings close to those 
given by dust maps (\citealt{sfd98}, \citealt{sf11}) and spectroscopically
derived metallicities (\citealt{cbg09}).  These particular investigations also
used fully consistent evolutionary tracks for the core He-burning phase to
generate synthetic HBs for direct comparisons with observed HBs and/or to
predict the periods of the $ab$- and $c$-type RR Lyrae variables in the target
GCs using the latest theoretical calibrations for $\log\,P_{ab}$ and $\log\,P_c$
as a function of luminosity, mass, $\teff$, and metallicity (\citealt{mcb15}).
Encouragingly, the models usually reproduced the observed periods to within
the uncertainties of the factors upon which periods depend, though not in the
case of M$\,$13 (see \citealt{dvk17}).  However, it would be possible to explain
the periods of its $c$-type variables if a reduced distance and/or an increased
reddening were adopted.  MS-fits to local subdwarfs were undertaken in some
studies as well (e.g., \citealt{vd18}); they also supported the ZAHB-based
distance scale.

Not only does such work give one added confidence in the ages that are derived,
but the resultant overlays of the isochrones onto the observed CMDs also provide
meaningful comparisons of the predicted morphologies of the isochrones with
those observed.  While small zero-point shifts to the synthetic colours
(typically $\lta 0.01$--0.02 mag) have to be applied to the models in order to
match the observed TOs, such offsets are not necessarily an indication of
deficiencies in the isochrones as they could easily be due to errors in the
photometric zero-points or to the assumption of a reddening that is somewhat
too large or too small or metal abundances that are too high or too low by
$\sim \pm 0.1$ dex.  The ACS observations originally obtained by \citet{sbc07}
provide a good example in support of this assertion.  When VBLC13
compared their isochrones to these observations, they generally had to shift the
models by $\delta(M_{F606W}-M_{F814W})_0 \sim -0.015$ mag in order to match the
TO colours.  No such offsets, or at least ones that amount to no more than a
a few thousandths of a magnitude, are needed when the same isochrones are
applied to the new reduction and calibration of the same data by \citet{nlp18}.
In fact, a favourable outcome of the 2018 calibration is that the fits of V-R
isochrones to $F606W,\,F814W$ observations are now much more consistent with
similar fits to $VI_C$ observations (see Fig.~\ref{fig:f5}).

Regardless, it should be appreciated that small colour offsets have no impact
on the inferred ages.  Rather, such offsets {\it must} be applied to match the
TO colours in order to derive the best estimate of the cluster age --- because
it is the superposition of the isochrones onto the arc of stars in the vicinity
of the TO that identifies which one has the same TO as the cluster;
for a discussion of this point, see VBLC13.  Isochrones from different sources
must be similarly registered to each other, as demonstrated in Fig.~\ref{fig:f4}
to ascertain whether the models for the same age and chemical abundances predict
the same TO luminosity.

In what follows, V-R isochrones are compared with recent IR observations of GCs
to assess their reliability in the regime of very low masses.  The next
subsection examines how well V-R isochrones match the temperatures and
selected colours of nearby subdwarfs with accurate distances that are located
along the upper MS and near the TO.  Finally, the constraints provided by a
recent empirical calibration of the absolute $I_C$ magnitude at the tip of
the RGB (TRGB) as a function of $(V-I_C)_0$ are studied. 

\begin{figure*}
\begin{center}
\includegraphics[width=\textwidth]{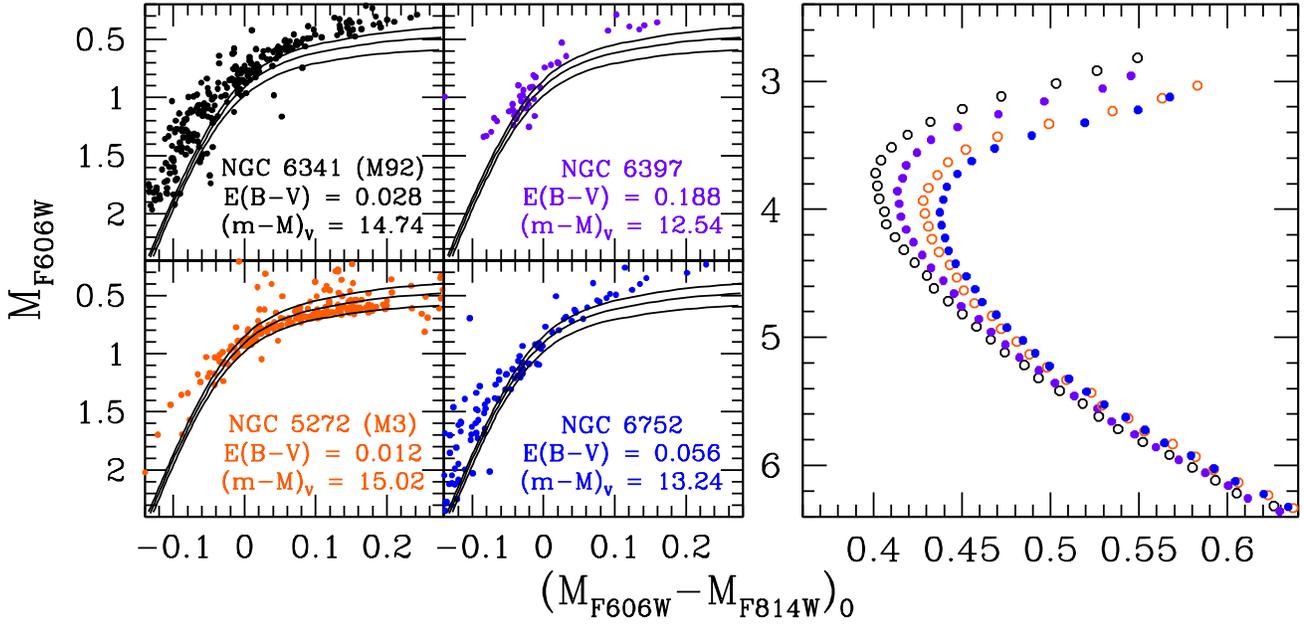}
\caption{{\it Left-hand panels:} Fits of the HB populations of 4 GCs, as
identified in each panel, to ZAHB loci for [Fe/H] $= -2.35, -2.0$, and $-1.5$
(in the order of increasing $M_{F606W}$ at a fixed colour) on the assumption of
the indicated values of $E(B-V)$ and $(m-M)_V$.  The ZAHBs assume $Y = 0.25$,
[O/Fe] $= +0.6$, and [$m$/Fe] $= +0.4$ for the other $\alpha$ elements.
{\it Right-hand panel} Comparison of the locations of the median fiducial
sequences for the upper MS, TO, and SGB stars in the same 4 clusters.  As in
the left-hand panels, the black, purple, orange, and blue points represent
M$\,$92, NGC$\,$6397, M$\,$3, and NGC$\,$6752, respectively.}
\label{fig:f6}
\end{center}
\end{figure*} 

\subsection{IR Photometry}
\label{subsec:ir}

The LMS portions of IR CMDs have a distinctive appearance insofar as such
colours as $J-K_S$ or $m_{F110W}-m_{F160W}$ remain nearly constant or become
bluer as the masses of the MS stars fall below $\sim 0.45\,\msol$, thereby
producing a knee-like morphology.  Because the absolute magnitude of this
feature is age-independent, it can be used to constrain GC ages, as first
pointed out by \citet{cbs09} and subsequently investigated by many researchers
(e.g., \citealt{bsv10}, \citealt{mtb15}, \citealt{cgk16}, \citealt{sdf18}).
However, aside from the concern raised by Saracino et al.~that BaSTI, DSEP,
and V-R isochrones predict absolute magnitudes of the MS-knee that differ by
a few tenths of a mag, which calls into question the reliability of this
reference point, more recent work has shown that the MS-knee is complicated
by the presence of multiple stellar populations.  This was revealed in the
truly spectacular IR CMD of NGC$\,$6752 that was obtained by \citet[see their
Fig.~1]{mmb19}.  Whereas the upper MS in this system was found to be very narrow
and well defined, the width of the LMS increased dramatically with increasing
magnitude beginning at the location of the MS knee.  In fact, such CMDs appear
to be quite typical of GCs; see the survey results by \citet{dmr22}.

\begin{figure*}
\begin{center}
\includegraphics[width=\textwidth]{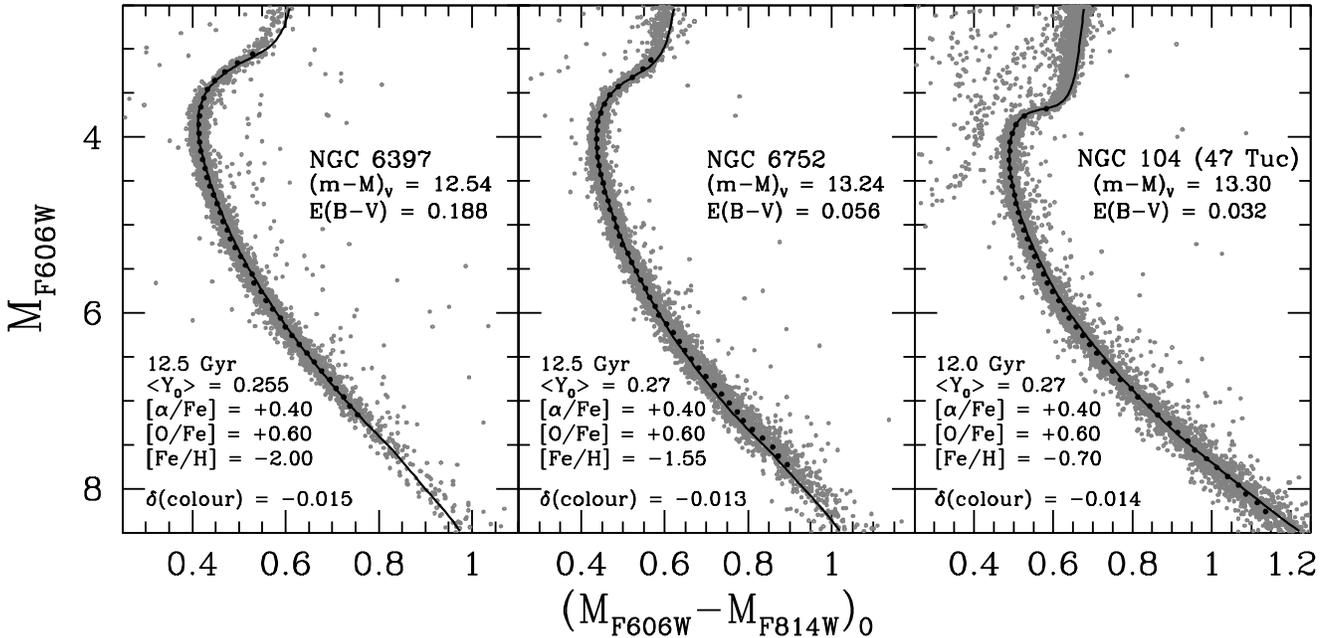}
\caption{Fits of isochrones for the indicated ages and chemical abundances
to the MS and TO portions of the NGC$\,$6397, NGC$\,$6752, and 47 Tuc, on the
assumption of the same values of $E(B-V)$ and $(m-M)_V$ that were adopted in
the previous figure.  The photometry is from \citet{sbc07}.  The small filled
circles in black reprsent the median fiducial sequences.}
\label{fig:f7}
\end{center}
\end{figure*} 

\citet{mmb19} attributed the observed spread in the colours of the LMS stars
of NGC$\,$6752 below the MS knee to star-to-star variations in the abundance of
oxygen.  Since isochrones have been generated for several O-enhanced mixtures
(e.g., {\tt a4xO\_p2}, {\tt a4xO\_p4}) as well as one mix with a very low O
abundance ({\tt a4ONN}), it is of considerable interest to find out how well
those models are able to explain the IR CMDs of GCs.  The primary goal here is
to overlay isochrones for the different mixtures of the metals onto the CMDs of
a few clusters in order to test the most basic predictions of the models.
More detailed studies, especially of the superb photometry obtained by
\citet{dmr22}, is well outside the scope of this particular paper.   To have
a first look at isochrones that span a wide range in metallicity, the decision
was made to examine $F110W,\,F160W$ observations of NGC$\,$6397
(\citealt{cgk18}), NGC$\,$6752 (\citealt{mmb19}, \citealt{dmr22}), and 47 Tuc
(\citealt{cgk16}), which have [Fe/H] values ranging from $\approx -2.0$ to
$\approx -0.7$.

To constrain the distances of NGC$\,$6397 and NGC$\,$6752, their HB populations
have been fitted to ZAHBs for $Y = 0.25$ and [Fe/H] $= -2.0$ and $-1.5$,
respectively; see the left-hand panels of Figure~\ref{fig:f6}, which also
illustrate how well the lower boundaries of the distributions of HB stars in
M$\,$92 and M$\,$3 can be matched by the appropriate ZAHBs for these two
GCs.  Because VBLC13 was able to obtain rather good fits to the \citet{sbc07}
observations of the cluster HB stars --- without any apparent difficulties
whatsoever --- using ZAHBs that differ only slightly from those computed for
this project, the $F606W,\,F814W$ data that appear in Fig.~\ref{fig:f6} have
been taken from the same source.  Indeed, the VBLC13 study left the impression
that the Sarajedini et al.~observations are quite homogeneous (as expected).
For instance, even though the isochrones required small zero-point offsets to
match the observed TO colours, these offsets were usually in the range from
$-0.01$ to $-0.025$ mag, independently of the cluster metallicity.
Cluster-to-cluster differences in applied colour shifts could easily be due in
part to small errors in the adopted reddenings or metal abundances. 

As noted in the introductory remarks of this section, there is considerable
evidence in support of the ZAHB-based distances that are obtained when
metallicities given by \citet{cbg09} and reddenings close to those derived
from the \citet{sfd98} dust maps are adopted (also see VBLC13).  To be sure,
without additional constraints, the uncertainties are very large in the case
of clusters that only have very blue HBs, such as NGC$\,$6397 and NGC$\,$6752,
since small differences in the adopted reddenings can have a huge impact on the
distance modulii that are derived from such fits (see Fig.~\ref{fig:f6}).
However, the assumed values of $E(B-V)$ and $(m-M)_V$ must be such that the main
sequences of the four clusters that are considered in Fig.~\ref{fig:f6} are
located relative to one another approximately as shown in the right-hand panel
in order to be consistent with the differences in their metallicities. 

At low metal abundances, neither the model $\teff$s nor the predicted
$(M_{F606W}-M_{F814W})_0$ colours along the MS are very dependent on [Fe/H].
In fact, the horizontal separation between the loci for M$\,$92 and M$\,$3 at
at $M_{F606W} \gta 5.5$ in the right-hand panel is consistent with the
predictions of the V-R isochrones in a differential sense.  This was
accomlished by adopting reddenings from the \citet{sfd98} dust maps for M$\,$3.
NGC$\,$6397, and NGC$\,$6752 (the values of $E(B-V)$ that are specified in the
left-hand panels of Fig.~\ref{fig:f6}), along with $E(B-V) = 0.028$ for M$\,$92,
which is higher than the dust map value by only 0.006 mag.  These reddenings,
coupled with the apparent distance moduli from the fits of the cluster HB
populations to the relevant ZAHBs resulted in the very agreeable comparison
of the cluster fiducials in the right-hand panel. 

One might question whether ZAHBs for $Y = 0.25$ are relevant to GCs with HBs
that are entirely to the blue of the instability strip, but the spectroscopic
study by \citet{vpg09} found that the HB stars in NGC$\,$6752 with $8500 <
\teff < 11,500$~K (near the top of the blue tail, where gravitational settling
is not effective) have close to the primordial He abundance.  Thus, it seems
to be a safe assumption that at least some of reddest HB stars in globular
clusters have $Y \approx 0.25$ (also see the study of M$\,$4 by \citealt{vgp12}). 
This is, anyway, of little concern as modest variations in $Y$ (or [Fe/H]; see
Fig.~\ref{fig:f6}) do not affect the location of the blue HB tail in optical
CMDs very much. 

In the case of 47 Tuc, $(m-M)_V = 13.30$ has been adopted so to be within 0.03
mag of the values found from fits of a suitable ZAHB and synthetic HB to
the observed HB stars (\citealt{dvk17}), from the eclipsing binary V69 by
\citet{bvb17}, and from {\it Gaia} DR2 parallaxes (\citealt{crc18}). 
\citet{tud20} also analyzed V69, as well as a second eclipsing binary, E32,
from which they obtained $(m-M)_0 = 13.29$, which is 0.08 mag larger than
the determination by Brogaard et al.  The difference in these results seems
to be mostly due to the adoption of different $\teff$s for the binary
components, according to K.~Brogaard (private communication); consequently,
it would be reasonable to adopt the average distance modulus from these two
investigations as the best estimate from the 47 Tuc binaries.  Nevertheless,
consistency with the ZAHB-based distances that have been adopted for the
more metal-deficient GCs would favour an apparent modulus near 13.30.  As for
the other clusters, the reddening from the \citet{sfd98} dust maps, $E(B-V) =
0.032$, has been adopted for 47 Tuc.

It is worth pointing out that the derived cluster distances agree rather well
with those reported by \citet{bv21}, who did an exhaustive literature review of
distance modulus determinations for most of the Galactic GCs, as well as
employing a number of different methods to obtain additional estimates of
$(m-M)_0$.  If the adopted apparent moduli are converted to $(m-M)_0$, on the
assumption of the aforementioned reddenings, one obtains 14.65 (14.65) for
M$\,$92, where the number enclosed by parentheses is from the Baumgardt \&
Vasiliev study, 11.96 (11.97) for NGC$\,$6397, 13.12 (13.08) for NGC$\,$6752,
14.98 (15.04) for M$\,$3, and 13.21 (13.28) for 47 Tuc.  Recall that the
measurement of the direct trigonometric parallax of NGC$\,$6397 by \citet{bcs18}
yielded $(m-M)_0 = 11.89 \pm 0.07$ for this system., which is consistent with
the other determinations to within $\approx 1\,\sigma$.  Clearly, the use of
ZAHBs to derive cluster distances is very competitive with other methods, even
when dealing with very blue HB populations.

\begin{figure*}
\begin{center}
\includegraphics[width=\textwidth]{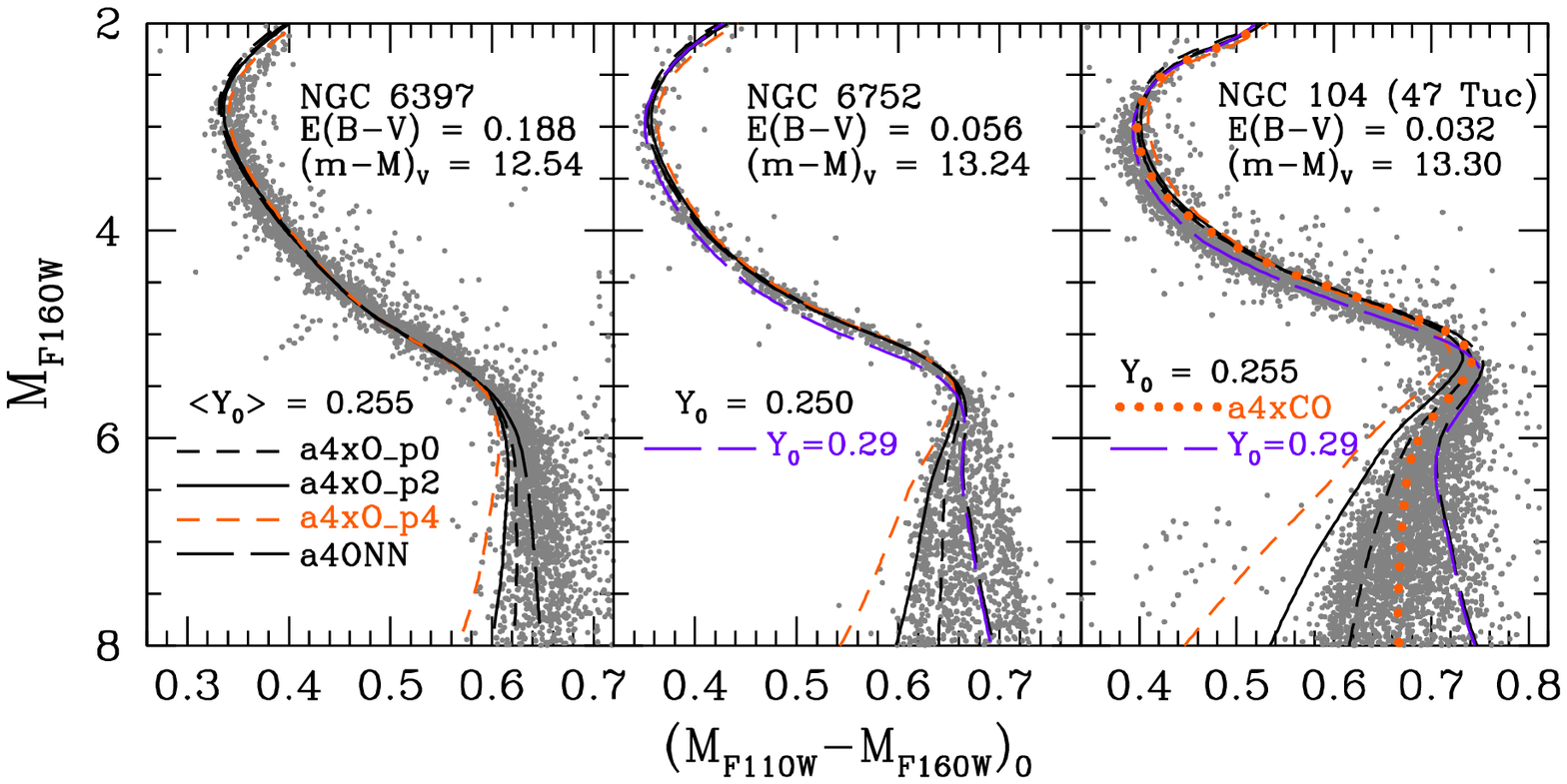}
\caption{Overlays of isochrones for several different mixtures of the metals
onto the IR CMDs of NGC$\,$6397 (\citealt{cgk18}), NGC$\,$6752 (\citealt{mmb19},
\citealt{dmr22}), and 47 Tuc (\citealt{cgk16}), on the assumption of
the indicated values of $E(B-V)$ and $(m-M)_V$.  These parameters and the
adopted ages are the same as in the previous figure.  The C, N, and O abundances
of the different mixtures are given in Table~\ref{tab:t1}.  In the case of
NGC$\,$6397, all of isochrones assume $\langle Y_0\rangle = 0.255$ while the 
isochrones that appear in the middle and right-hand panels, which have been
generated for the same mixtures that are identified in the left-hand panel,
assume $Y_0 = 0.250$ and $Y_0 = 0.255$, respectively.  Additional isochrones
for the {\tt a4ONN} mix but assuming $Y_0 = 0.29$ have been superimposed on the
NGC$\,$6752 and 47 Tuc CMDs (the long-dashed curves in purple).  An isochrone
for the {\tt a4xCO} mix (orange filled circles) has also been plotted, but only
in the right-hand panel.}
\label{fig:f8}
\end{center}
\end{figure*} 

Before turning to the IR CMDs, it is worthwhile to show that the V-R isochrones
are able to provide exceedingly good fits to the MS populations of NGC$\,$6397,
NGC$\,$6752, and 47 Tuc on CMDs constructed from the $F606W,\,F814$ filters.
This is illustrated in Figure~\ref{fig:f7}, which shows that the stellar
models accurately reproduce the observed morphologies of the cluster CMDs from
the SGB down to $M_{F606W} \lta 8.0$.  Although small colour offsets had to be
applied to the models to achieve the results that are shown, they are of no
consequence.  As already mentioned, significantly smaller adjustments are
required to obtain comparable fits of the same isochrones to the updated 
versions of the same CMDs by \citet{nlp18}.  Regardless, there are bound to
be small differences in the zero points of the observed and synthetic 
photometry, small errors in the assumed cluster properties, and errors that
are difficult to evaluate in the stellar models due to inadequacies in the
treatment of convection, the atmospheric boundary condition, etc.  In fact,
one would not have expected the agreement between theory and observations
to be anywhere near as good as in the examples shown in Fig.~\ref{fig:f7}.

It should be appreciated that the isochrones have effectively been {\it overlaid}
onto the observed CMDs (aside from the application of small colour shifts) on
the assumption of well supported estimates of the cluster properties that are
independent of the isochrone fits to the MS and TO stars.  The basis for the
adopted values of $E(B-V)$ and $(m-M)_V$ has already been explained.  The
metallicities of NGC$\,$6397 and NGC$\,$6752 are within 0.01 dex of the
spectroscopic values given by \citet{cbg09}, while [Fe/H] $= -0.70$ has been
assumed for 47 Tuc because it seems to be difficult to satisfy the constraints
from its eclipsing binaries if it has a lower metallicity (see \citealt{bvb17},
Paper II).  The evidence in support of Pop.~II stars having [O/Fe] $= +0.6$ and
[$m$/Fe] $= +0.4$ for the other $\alpha$ elements being was provided in
\S~\ref{sec:models}.  Insofar as the mean He abundances are concerned:
variations amounting to $\Delta\,Y \sim 0.01$ appear to be present in
NGC$\,$6397 (\citealt{mmp12}), while they are apparently closer to $\sim 0.03$
in both NGC$\,$6752 (\citealt{mmp13}, \citealt{dfc15}) and 47 Tuc
(\citealt{scp16}, \citealt{dvk17}). Accordingly, $\langle Y_0\rangle = 0.255$
should be reasonably close to the mean value in NGC$\,$6397, while $\langle
Y_0\rangle = 0.27$ should be a good estimate for NGC$\,$6752 and 47 Tuc.  Given
that there is little ambiguity in the fits of the selected isochrones to the TO
stars, the inferred cluster ages should be quite accurate, subject to the
reliability of the adopted cluster properties.

The main results of this section are shown in Figure~\ref{fig:f8}, which 
superimposes isochrones for the mixtures of the metals that are identified in
the left-hand panel onto IR observations of NGC$\,$6397, NGC$\,$6752, and 47
Tuc.  Since the star-to-star variations of $Y$ are small in NGC$\,$6397, all of
the isochrones that have been compared with its CMD assume the same He abundance
($\langle Y_0\rangle = 0.255$). To illustrate the effects of $\Delta\,Y =
0.03$--0.04 in the other two GCs, isochrones for just the {\tt a4ONN} mix have
been plotted for $Y = 0.29$ as well as for $Y_0 = 0.25$ in the middle and
right-hand panels. (The minimum value of $Y_0$
is expected to be somewhat larger in 47 Tuc than in NGC$\,$6752 due to Galactic
chemical evolution.)  Had isochrones for $Y = 0.29$ been plotted for other
mixtures (i.e., {\tt a4xO\_p0}, {\tt a4xO\_p2}, and {\tt a4xO\_p4}), they would
have coincided with the long-dashed curve in purple above the knee, and been
nearly coincident with the loci for the same mixtures and the low value of $Y_0$
below the knee.  The location of an isochrone for the {\tt a4xCO} mix, assuming
$Y_0 = 0.255$, is represented by small filled circles in orange in only the
right-hand panel.  Note that the {\tt a4xO\_p2} and {\tt a4ONN} mixtures have
the same value of [CNO/Fe]; hence, the spread in LMS colours between the
respective isochrones (the solid and long-dashed loci) is expected for stars of
the same C$+$N$+$O abundance but with variations in the ratio C:N:O that result
from CN- and ON-cycling.  A few isochrones for very low C and O and high N with
the same [CNO/Fe] as the {\tt a4xO\_p0} mixture were presented in Paper I.  As
illustrated in Fig.~12 of that investigation, they superimpose the isochrones
for the {\tt a4ONN} mixture at the same [Fe/H] almost exactly.

These few remarks and the information provided in the figure
caption should provide a sufficiently detailed description of what has been 
plotted; but before assessing the performance of the stellar models, some of
the features of the isochrones should be highlighted.  First, at a fixed value
of $Y$, the isochrones for all of the metal abundance mixtures that have been
considered overlay one another above the MS knee, but they are widely separated
below the knee.  Varying $Y$ has the opposite effect; predicted
$(M_{F110W}-M_{F160W})_0$ colours are dependent on the He abundance 
only above the knee.  Second, the morphologies of the isochones are fairly
sensitive functions of both [Fe/H] and the mixture of the metals.  Below the
knee, isochrones are nearly vertical at low metallicities and at low oxygen
abundances, but they bend strongly back to the blue at high [Fe/H] values and/or
high O abundances.  Third, the $M_{F160W}$ magnitude of the knee becomes
progressively brighter with increasing [Fe/H] (compare the location of the knee
in the three panels of Fig.~\ref{fig:f8}) and with increasing [O/Fe] at a fixed
metallicity, which is most evident in the third panel, but also apparent in the
other panels if magnified versions of the figure are viewed.

As regards the capabilities of the stellar models in reproducing IR CMDs, their
most obvious deficiency is that they are unable to explain the reddest stars
below the knee.  The bluest stars do not present a problem and, indeed, it is
encouraging that the models indicate a preference for [O/Fe] $= +0.6$ at low
metallicities and a lower value by 0.1 dex or so at [Fe/H] $\approx -0.70$,
which is very similar to the observed trend in nearby Pop.~II stars (see, e.g.,
\citealt{ncc14}).  The IR observations thus provide a further justification for
such O abundances in addition to those given in \S~\ref{sec:models}.  Higher N
might help to reduce the discrepancies, though the {\tt a4ONN} mix assumes
[N/Fe] $= +1.48$ (along with very low C and O, see Table~\ref{tab:t1}), which
is close to the maximum value found in spectroscopic stuides of GCs (e.g.,
\citealt{bcs04}, \citealt{cbs05}).  It seems more likely that either the reddest
LMS stars have chemical abundances that are quite different from those assumed
in any of the mixtures considered in this study or that the MARCS models do not
treat the molecular sources of opacity well enough in cool, LMS stars.
\citet{mmb19} had better success reproducing the spread in the observed colours
below the knee in NGC$\,$6752 (see their Fig.~3), but they did not allow for the
effects of different O abundances on the $\teff$s of low-mass stellar models
(because of the lack of suitable isochrones at the time).  It would be
interesting to find out how much their results would change if they allowed for
the effects of O abundance variations on $\teff$s of stars.

Other than this, the fits of the stellar models to the NGC$\,$6752 CMD look
quite good; even the width of the MS above the knee is well matched by
isochrones for $Y_0 = 0.25$ and 0.29.  In a systematic sense, the CMD of 47 Tuc
is also in satisfactory agreement with the model predictions, though there seems
to be a zero point offset between them.  There is, in fact, some evidence in
support of this possibility.  As part of the same survey that resulted in the
47 Tuc observations, \citet{cgk16} obtained an IR CMD for NGC$\,$6752.  An
examination of those data revealed that the median fiducial points, which are
tabulated in their paper, are in superb agreement with the CMD of NGC$\,$6752
in Fig.~\ref{fig:f8} above the knee, and with the isochrones for the
{\tt a4xO\_p0} mix below the knee, {\it if} the colours are adjusted to the red
by 0.01--0.015 mag.  If the same zero-point offset was applied to their data
for 47 Tuc, the isochrones would provide an improved fit to those data.

The IR CMD of NGC$\,$6397 appears to be the most problematic one for the 
present isochrones as the models do not reproduce the shape of the observed
CMD above the knee very well.  However, some improvement is obtained if
isochrones for [Fe/H] $= -1.9$ are used in the comparison instead of those for
[Fe/H] $= -2.0$.  The same conclusion was reached by \citet{cgk18}, who fitted
V-R isochrones from \citet{vbf14} for [$\alpha$/Fe] $= 0.4$ (i.e., for the same
metal abundances as in the {\tt a4xO\_p0} mix) to their photometry.  This shows
that IR observations can be used to set limits on the overall metalllicities of
GCs.  In any case, even though significant advances have been made in the
modelling of such data, it is clear that much remains to be done.

\subsection{Field Halo Stars}
\label{subsec:halo}

\begin{figure*}
\begin{center}
\includegraphics[width=\textwidth]{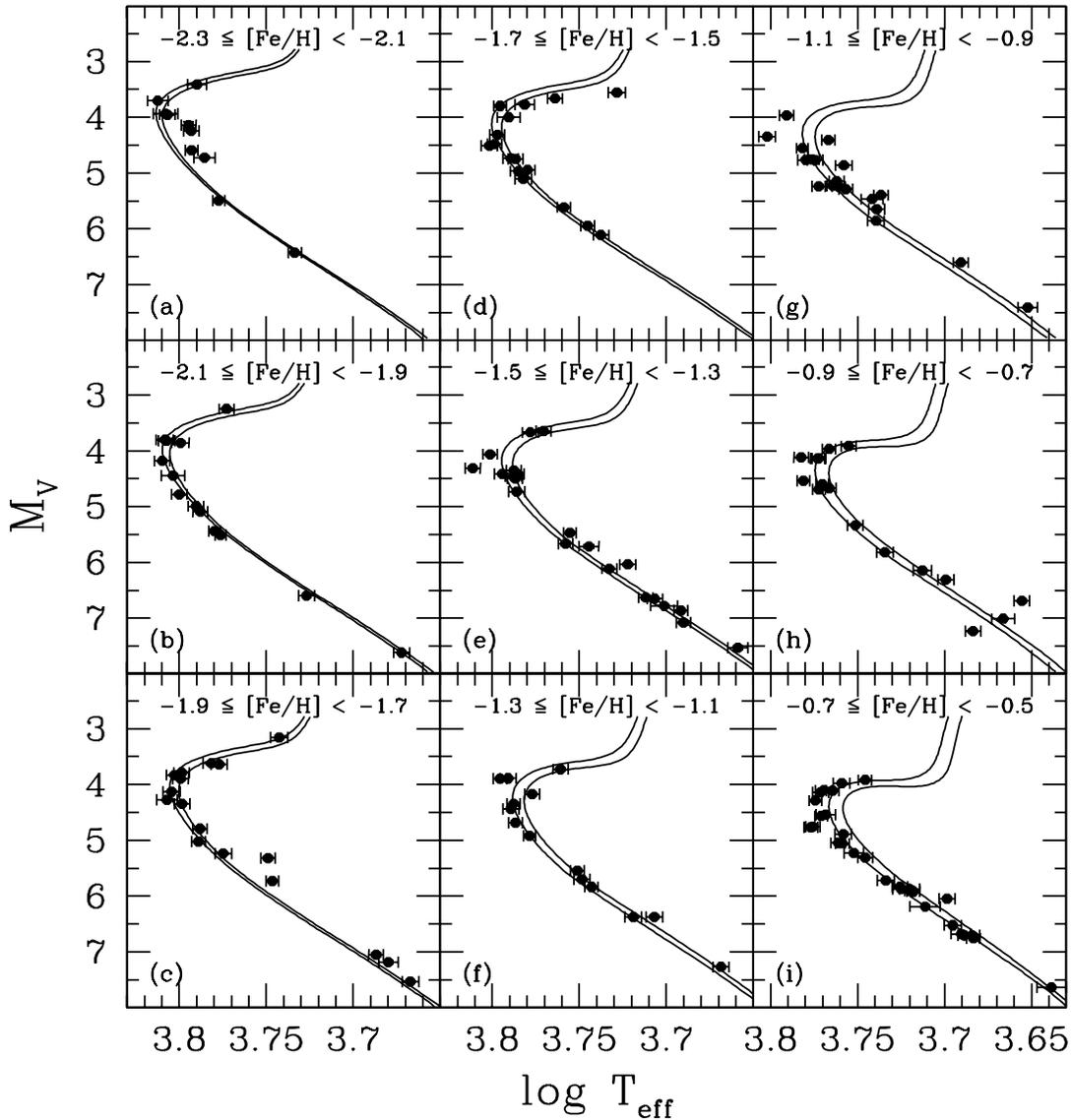}
\caption{Overlays of nearby subdwarfs with [Fe/H] values within the indicated
ranges by 12.5 Gyr V-R isochrones for $Y = 0.25$, [O/Fe] $= +0.6$, [$m$/Fe] $=
+0.4$ for the other $\alpha$ elements, and metallicities given by the upper and
lower limits of those ranges.  The temperatures and their uncertainties are from
the study by \citet{crm10}.  The absolute visual magnitudes and their
uncertainties, which are smaller than the sizes of the filled circles that are
used to represent the stars, were derived from EDR3 parallaxes (see the text
for details).}
\label{fig:f9}
\end{center}
\end{figure*} 

Nearby subdwarfs with accurate distances provide one of the primary constraints
on the temperatures and colours of stellar models, and they have frequently
been used to check the reliability of Victoria computations and the isochrones
derived from them (see, in particular, \citealt{vcs10}, \citealt{vbf14}).  Such
studies have shown shown, for instance, that V-R isochrones reproduce the
locations of solar neighbourhood Pop.~II stars on the
$(\log\,\teff,\,M_V)$-diagram to well within the $\sim 70$~K uncertainties of
the temperatures derived for them from the application of the IRFM by
\citet{crm10} when their absolute visual magnitudes are based on {\it Hipparcos}
parallaxes.  Furthermore, the same models are able to reproduce the observed
$(V-K_S)_0$ and $(V-I_C)_0$ colours (usually to within 0.0-0.02 mag) if the
colour transformations based on MARCS model atmospheres and synthetic spectra
are adopted.

Not surprisingly, the huge improvement to the parallaxes of nearby stars
resulting from {\it Gaia} observations, and the consequent large increase in the
number of stars with very accurate distances, has only served to confirm such
findings.  To demonstrate this, {\it Gaia} EDR3 parallaxes (\citealt{gai18})
have been obtained from the
{\it Gaia} archive.\footnote{https://gea.esac.esa.int/archive} for a large
fraction of the metal-poor stars in the \citet{crm10} catalogue.  These
parallaxes were corrected for systematic errors as a function of magnitude,
colour, and ecliptic latitude using the recipe given by \citet{lbb21}, even
though such corrections are inconsequential for stars with parallaxes larger
than a few {\it mas}, as in the case of the selected subdwarfs.  Because
there are over 200 MS and TO stars in the adopted data set, they could be
separated into bins spanning 0.2 dex in [Fe/H], assuming the metallicities
given by \citet{crm10}, who also provide effective temperatures and their
uncertainties.  Stars known to be members of binaries or to show evidence of
variability were not included in the sample.

\begin{figure*}
\begin{center}
\includegraphics[width=\textwidth]{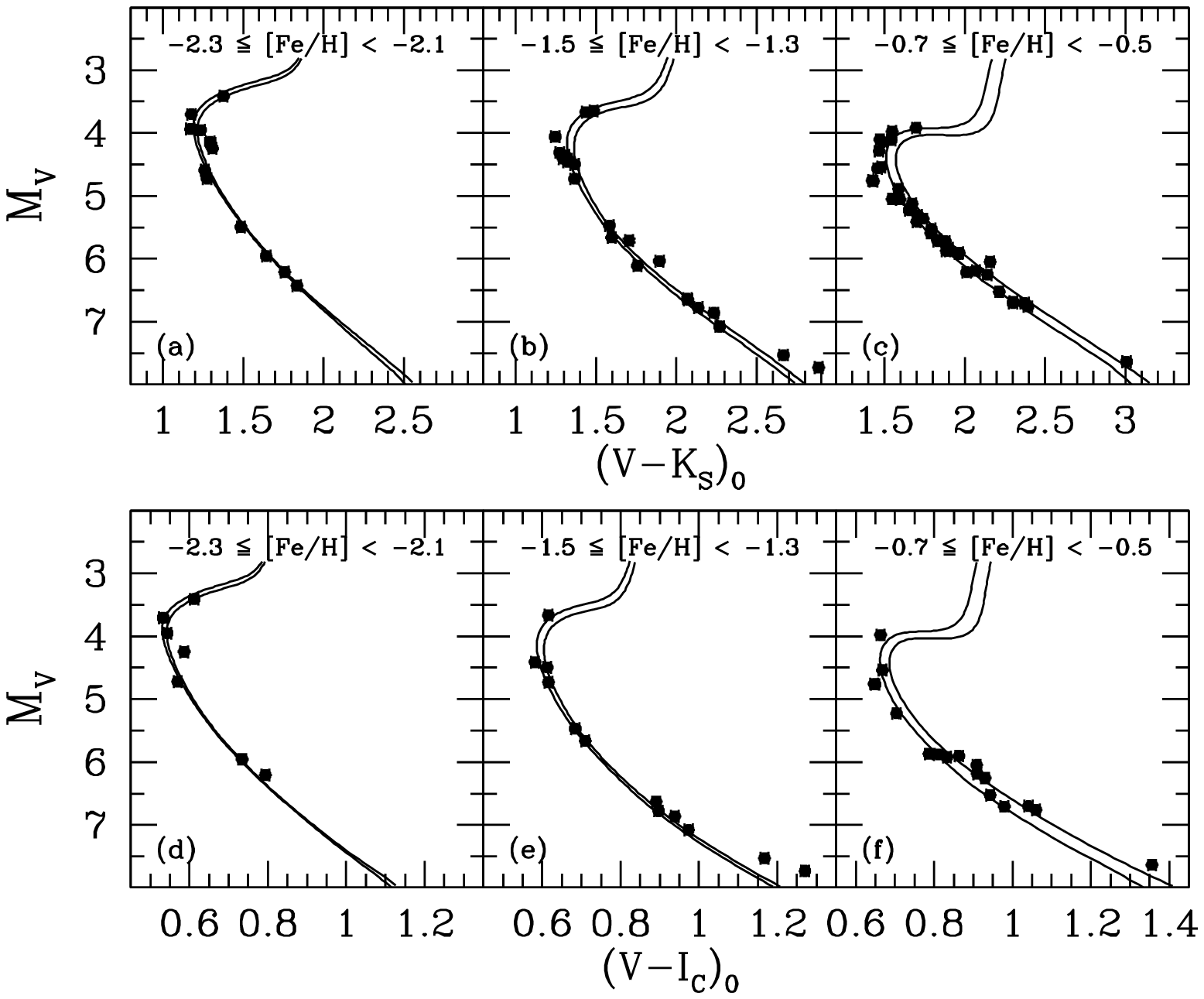}
\caption{As in the previous figure, except that the comparisons with the
isochrones involve $V-K_S$ colours (the top row of panels) or $V-I_C$ colours
(the bottom row of panels).  Some of the stars that appear in panels (d)--(f)
were taken from the investigations by \citet{ogc17} and \citet{cmo17} given
that $V-I_C$ colours are provided for only a small number of subdwarfs in the
\citet{crm10} data set.  The comparisons have been limited to the metallicity
bins along the main diagonal in Fig.~\ref{fig:f9} that runs from the upper
left-hand corner to the lower right-hand corner, in order to reduce the space
taken by the plots and still sample the full range of metallicity from $-2.30$
to $-0.5$.  The reddenings of the subdwarfs were derived from the 3D maps given
by \citet{clv17}.}
\label{fig:f10}
\end{center}
\end{figure*} 

Plots of $M_V$ as a function of $\log\,\teff$ for the binned subdwarfs are
shown in Figure~\ref{fig:f9}.  For instance, panel (a) in the top left-hand
corner contains the stars with $-2.3 \le$ [Fe/H] $< -2.1$ along with 12 Gyr
V-R isochrones for [Fe/H] $= -2.3$ and $-2.1$.  All of the isochrones in this
figure assume $Y = 0.25$, [O/Fe] $= +0.6$, and [$m$/Fe] $= +0.4$ for all of the
other $\alpha$ elements.  For the most part, the isochrones provide good fits
to the subdwarfs, though there are a few stars in this bin that are redder
than the stellar models at $M_V \sim 4.5$.  In the next metallicity bin (see
panel b), the stars define a much tighter sequence and their locations on
the $(\log\,\teff,\,M_V)$-diagram are reproduced particularly well by the
relevant isochrones.  Although there are a few outliers in most of the panels,
the MS and TO portions of the isochrones that are plotted in each bin are
approximately where they should be based on the subdwarfs that have been 
plotted.  The discrepant stars could well be unrecognized binary components
or there may be some issues with their metallicities and/or temperatures. 
Most of the TO stars and subgiants are reasonably well fitted, suggesting
that they have ages near 12.5 Gyr, though it would be worthwhile to fit
isochrones directly to the subgiants in order to obtain best estimates of their
ages for comparisons with the ages of the Galactic GCs.  The most metal-rich
stars (see panel i) do appear to be younger than 12.5 Gyr as the TO stars are
somewhat brighter and bluer than the isochrones that have been plotted.

Figure~\ref{fig:f10} is similar to the previous figure, except that the
comparisons of the subdwarfs with the isochrones involve $V-K_S$ colours (the
top row of panels) or $V-I_C$ colours (the bottom row of panels).  As
\citet{crm10} tabulate $JHK_S$ magnitudes for all of the stars, but $BVRI_C$
magnitudes for only a relatively small subset of them, a few additional stars
with $V-I_C$ colours were drawn from the investigations by \citet{ogc17} and
\citet{cmo17}.  In order to reduce the space taken by these plots and still
sample the full range of metallicity from $-2.30$ to $-0.5$, only the  
metallicity bins along the main diagonal in Fig.~\ref{fig:f9} that runs from
the upper left-hand corner to the lower right-hand corner are shown.  Indeed,
they are sufficient to serve the purpose of demonstrating that V-R isochrones
reproduce the CMDs of local halo stars rather well.  Note that the reddenings
of the subdwarfs were derived from the 3D maps given by
\citet{clv17}.\footnote{https://stilism.obspm.fr}  For the few stars that are
subject to small amounts of reddening (most are unreddened), the $E(B-V)$
values tend to be quite close to those determined from analyses of the
interstellar Na I D-lines by, e.g., \citet[see their Table 1]{mcr10}.

A forthcoming paper will examine the implications of local subdwarfs for GC
distances, as derived from the MS-fitting method, and determine the ages of
the few subdwarfs with accurate distances that are located between the MSTO
and the RGB.

\subsection{TRGB Luminosities}
\label{subsec:trgb}

\citet[hereafter JL17]{jl17} have obtained some of the
most accurate calibrations to date of the
absolute magnitude of the tip of the RGB (TRGB) as a function of colour, based
on observations of several nearby galaxies with the zero-point determined from
photometry of NGC$\,$4258 and the LMC, which have precise geometric distances.  
Their relation for $M_I$ as a function of $(V-I)_0$, in the Johnson-Cousins
system, is represented by the solid curve in orange in the left-hand panel of
Figure~\ref{fig:f11} along with the associated $1\,\sigma$ uncertainties (the
orange dashed loci that are offset from the solid curve by $\pm 0.058$ mag at a
fixed colour).  Also shown are the estimated locations of TRGB stars in
$\omega$ Cen, M$\,$5, and 47 Tuc, though the $M_I$ values are not the same
as those reported by \citet[\citealt{bfp01}]{bfr04} for $\omega$ Cen and
47 Tuc or by \citet{vcs13} for M$\,$5.  Only the apparent magnitudes of the
TRGB, which includes corrections for under-sampling due to the limited number
of very bright RGB stars in GCs, have been taken from these sources.  According
to Bellazzini et al.~, the TRGB occurs in $\omega$ Cen and 47 Tuc at $I = 9.84$
and 9.40, respectively, whereas Viaux et al.~give $I^{\rm TRGB} = 10.27$ for
M$\,$5.

\begin{figure*}
\begin{center}
\includegraphics[width=\textwidth]{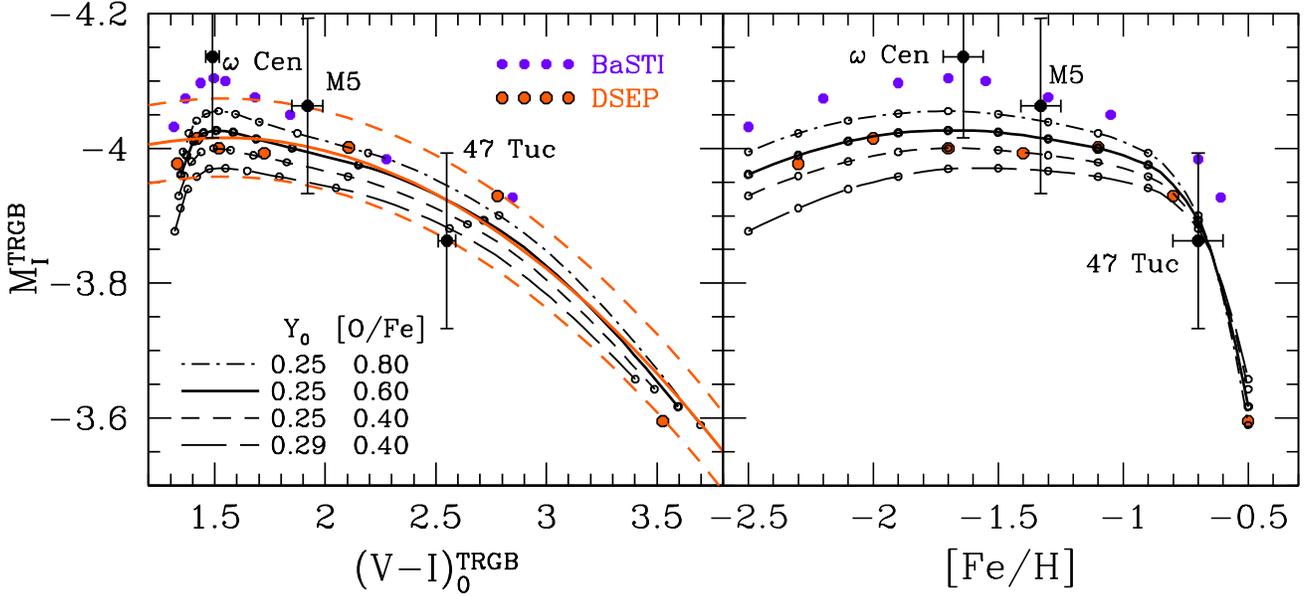}
\caption{{\it Left-hand panel:}~Plot of $M_I$ as a function of $V-I)_0$ at the
TRGB from JL17 (the solid curve in orange).  The parallel dashed loci
indicate the $1\,\sigma$ uncertainties associated with the solid curve.  The
absolute $I$ magnitudes of RGB tip stars in $\omega$ Cen, M$\,$5, and 47 Tuc
are represented by filled circles and the attached error bars (see the text).
The black curves, as identified in the lower left-hand corner of the plot,
were derived from V-R isochrones for the indicated initial He abundances
($Y_0$) and values of [O/Fe] for each of the metallicities from $-2.5$ to $-0.5$
(the small open circles in the direction from left to right).  Predictions from
the latest BaSTI and DSEP isochrones have been plotted by filled circles in 
purple and in orange, respectively.  {\it Right-hand panel:}~Similar to the
left-hand panel, except that the results from the various isochrones and for
the three GCs are plotted as a function of [Fe/H].}
\label{fig:f11}
\end{center}
\end{figure*}

\begin{figure*}
\begin{center}
\includegraphics[width=\textwidth]{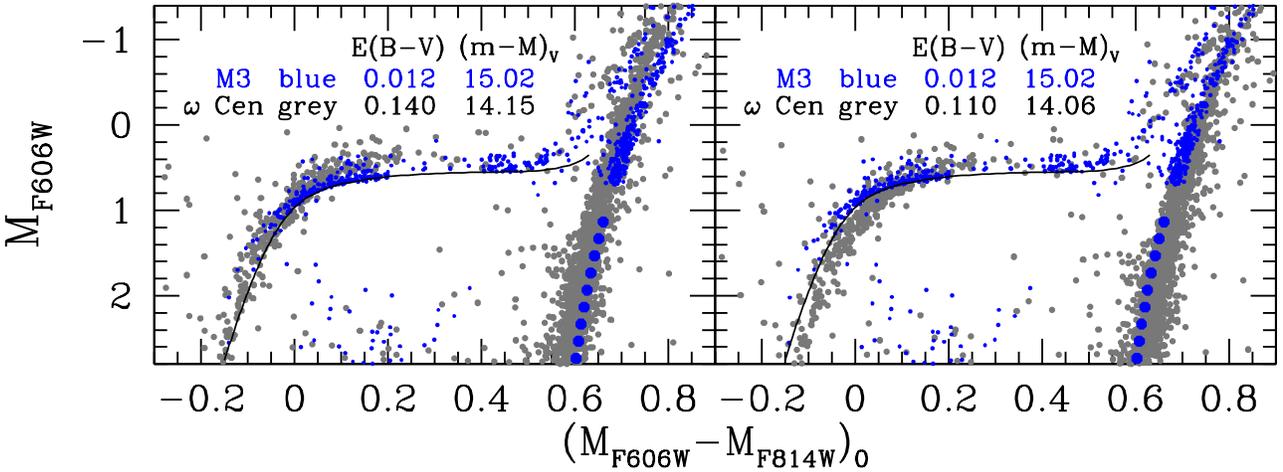}
\caption{Overlay of the CMDs for the HB and adjacent RGB populations in
M$\,$3 and $\omega$ Cen on the assumption of two different estimates of
$E(B-V)$ and $(m-M)_V$ for the latter.  The photometry is from \citet{sbc07}.
The large filled circles in blue define the median points along the lower RGB
of M$\,$3.  As indicated, the same properties of M$\,$3 are assumed in both
panels.  The solid curve in black represents the same ZAHB for [Fe/H] $= -1.50$
that was fitted to the M$\,$3 HB in Fig.~\ref{fig:f6}.} 
\label{fig:f12}
\end{center}
\end{figure*}

Because fits of Victoria ZAHB models to cluster HB populations yield distance
moduli that are comparable with current best estimates (e.g., \citealt{bv21}),
the aforementioned apparent magnitudes have been converted to $M_I$ values
assuming ZAHB-based determinations of $(m-M)_V$ and nominal $E(B-V)$ values
given by the \citet{sfd98} dust maps.  Although \citet{bfr04} argued that the
reddening of $\omega$ Cen is $E(B-V) = 0.11\pm 0.01$, it is simply not
possible to obtain a consistent superposition of the blue HB populations of
M$\,$3, which is nearly unreddened, and $\omega$ Cen if such a low reddening
is adopted for the latter.  This is illustrated in the right-hand panel of
Figure~\ref{fig:f12}.  If $\omega$ Cen has $E(B-V) = 0.11$ and $(m-M)_V =
14.06$, as derived by \citet{tkp01} from the application of an empirical IR
verus surface brightness relation by \citet{di98} to observations of the
eclipsing binary OGLEGC 17, its blue HB tail lies below both the M$\,$3 HB and
a ZAHB for [Fe/H] $= -1.50$.\footnote{As reported in their paper,
\citet{tkp01} found a much smaller distance modulus, $(m-M)_V = 13.78$, from
the measured bolometric luminosities of the binary components.}  The extended
blue tail of a ZAHB on optical CMDs is predicted to be nearly independent of
metallicity and, because it is almost vertical, to be an very good contraint on
the cluster reddening.  Since $\omega$ Cen has a mean metallicity of [Fe/H]
$= -1.64$ (\citealt{cbg09}), it is a further difficulty with the adopted
properties in the right-hand panel that the RGB of the more metal rich cluster,
M$\,$3, lies on the blue side of the $\omega$ Cen giant branch.

\begin{table*}
\centering
\caption{Properties of 12.5 Gyr Stellar Models at the RGB Tip$^{a}$}
 \label{tab:t2}
\smallskip
\begin{tabular}{cccccccccccc}
\hline
\hline
\noalign{\smallskip}
 [Fe/H] & ${\cal M}^{b}$ & $\log L/L_\odot$ & $\log\,\teff$ & $\log\,g$ &
   ${\cal M}_{\rm sh}^{c}$  & $Y_{\rm surf}$ & $M_V$ & $M_I^{d}$ & 
   $M_{F606W}^{e}$ & $M_{F814W}^{e}$ & $M_{K_S}$ \\
\noalign{\smallskip}
\hline
\noalign{\smallskip} 
\multispan{12} {\hfil mix: {\tt a4xO\_p2}, $Y_0 = 0.25$ \hfil} \\
\noalign{\vskip 3pt}
 $-2.50$ & 0.7897 & 3.2236 & 3.6414 & 0.6304 & 0.4998 & 0.2442 & $-2.6140$ &
   $-3.9611$ & $-2.9728$ & $-3.9829$ & $-5.6252$ \\
 $-2.30$ & 0.7910 & 3.2402 & 3.6362 & 0.5932 & 0.4974 & 0.2448 & $-2.6169$ &
   $-3.9895$ & $-2.9841$ & $-4.0118$ & $-5.7053$ \\
 $-2.10$ & 0.7930 & 3.2566 & 3.6294 & 0.5510 & 0.4948 & 0.2461 & $-2.6048$ &
   $-4.0103$ & $-2.9827$ & $-4.0329$ & $-5.7956$  \\
 $-1.90$ & 0.7963 & 3.2732 & 3.6207 & 0.5018 & 0.4923 & 0.2475 & $-2.5734$ &
   $-4.0229$ & $-2.9658$ & $-4.0460$ & $-5.8996$  \\
 $-1.70$ & 0.8019 & 3.2896 & 3.6104 & 0.4467 & 0.4900 & 0.2489 & $-2.5195$ &
   $-4.0267$ & $-2.9313$ & $-4.0503$ & $-6.0126$  \\
 $-1.50$ & 0.8094 & 3.3061 & 3.5989 & 0.3884 & 0.4877 & 0.2508 & $-2.4409$ &
   $-4.0239$ & $-2.8783$ & $-4.0478$ & $-6.1324$  \\
 $-1.30$ & 0.8220 & 3.3224 & 3.5861 & 0.3276 & 0.4856 & 0.2523 & $-2.3267$ &
   $-4.0141$ & $-2.7963$ & $-4.0374$ & $-6.2576$  \\
 $-1.10$ & 0.8404 & 3.3391 & 3.5719 & 0.2645 & 0.4835 & 0.2545 & $-2.1504$ &
   $-4.0004$ & $-2.6523$ & $-4.0201$ & $-6.3904$  \\
 $-0.90$ & 0.8673 & 3.3552 & 3.5566 & 0.2000 & 0.4815 & 0.2568 & $-1.8250$ &
   $-3.9755$ & $-2.3505$ & $-3.9871$ & $-6.5272$  \\
 $-0.70$ & 0.9042 & 3.3702 & 3.5408 & 0.1405 & 0.4795 & 0.2592 & $-1.1765$ &
   $-3.8934$ & $-1.7286$ & $-3.9008$ & $-6.6653$  \\
 $-0.50$ & 0.9508 & 3.3847 & 3.5246 & 0.0823 & 0.4773 & 0.2615 & $-0.0209$ &
   $-3.6169$ & $-0.6656$ & $-3.6473$ & $-6.8054$  \\
\noalign{\smallskip}
\hline
\noalign{\smallskip}
\end{tabular}
\begin{minipage}{1\textwidth}
$^{a}$~This and similar tables for the other He and metal abundance mixtures
 may be downloaded from the web site given in the Data Availability Section. \\
$^{b}$~Masses are specified in solar units ($\msol$).  Mass loss during RGB
 evolution has not been treated.\\
$^{c}$~The mass interior to the center of the H-burning shell, where the H
 abundance is one-half of the surface abundance. \\ 
$^{d}$~Predicted $I$ magnitudes are in the Johnson-Cousins photometric system. \\
$^{e}$~Predicted $F606W$ and $F814W$ magnitudes are in the {\it HST} Advanced
 Camera for Surveys (ACS) photometric system. \\
\phantom{~~~~~~~~~~~~~~~~~~~~~}
\end{minipage}
\end{table*}

All of these problems are resolved if $\omega$ Cen has $E(B-V) = 0.14$ from
the \citet{sfd98} dust maps, which corresponds to an actual $E(B-V)$ near 0.13
mag for the stellar populations that are portrayed in Fig.~\ref{fig:f12}, and
the distance modulus is obtained by matching a ZAHB to the bluest HB stars.
Encouragingly, \citet{bib19} derived almost exactly the same reddening and
distance modulus from their analysis of optical/near-IR observations of RR
Lyrae stars in $\omega$ Cen.  Insofar as the other two GCs are concerned: the
HB of M$\,$5, which extends well to the red of the instability strip, is
well reproduced by a ZAHB for [Fe/H] $= -1.33$ (\citealt{cbg09}) if $E(B-V) =
0.038$ (\citealt{sfd98}) and $(m-M)_V = 14.38$ (see VBLC13).  As already
discussed, the preferred parameters for 47 Tuc are $E(B-V) = 0.032$, from the
same dust maps, and $(m-M)_V = 13.30$. If these properties are adopted along
with $R_I = A_I/E(B-V) = 1.885$ from \citet{cv14}, it is a straightforward
exercise to convert the apparent $I^{\rm TRGB}$ magnitudes given above to the
values of $M_I^{\rm TRGB}$ that have been plotted in Fig.~\ref{fig:f11}.  The
uncertainties of $M_I$ are the same as those given by Bellazzini et al.~(2004,
2001) for $\omega$ Cen and 47 Tuc, and by \citet{vcs13} for M$\,$5.  The TRGB
magnitudes for the three GCs are clearly in good agreement with the JL17
relation to well within their respective $1\,\sigma$ uncertainties.  This
consistency can be viewed as further evidence in support of Victoria ZAHB
models.  For instance, the significantly larger true distance modulus that was
adopted by \citet{vcs13}, $(m-M)_0 = 14.43$, from their review of published
determinations, results in the considerably less satisfactory value of
$M_I^{\rm TRGB} = -4.17 \pm 0.13$ from the perspective of the JL17 results.

Perhaps somewhat fortuitously given the uncertainties, the TRGB predictions from
the V-R isochrones for the same O abundance that is favoured from other
considerations (i.e., [O/Fe] $= +0.6$) provides the closest match to the JL17
relation between $M_I^{\rm TRGB}$ and $(V-I)_0$; see the left-hand panel of
Fig.~\ref{fig:f11}.  Table~\ref{tab:t2} lists the properties of the stellar
models along this particular sequence, including predicted TRGB absolute
magnitudes for the Johnson-Cousins $VI$ filters, the {\it HST} ACS 
$F606W,\,F814W$ passbands, and the 2MASS $K_S$ filter.  Data files containing
the information in this table and for the other theoretical results that have 
been plotted, which illustrate the effects of varying $Y$ and [O/Fe], may be
downloaded from the web site that is specified in the Data Availability
section of this paper.

The left-hand panel of Fig.~\ref{fig:f11} also shows the TRGB $M_I$ versus
$(V-I)_0$ relations from BaSTI and DSEP computations.  The TRGB models were
taken to be the most luminous points along isochrones that were generated via
their respective web sites (see footnotes 5 and 6) for an age of 12.5 Gyr,
[$\alpha$/Fe] $= +0.4$, $-2.5 \le$ [Fe/H] $\le -0.5$, and initial He abundances
that vary with the mass-fraction abundance of the metals, $Z$, according to
an adopted enrichment ratio $\Delta\,Y/\Delta\,Z = 1.31$ (BaSTI) or $1.54$
(DSEP) with, in turn, $Y_p = 0.247$ and 0.245 for the primordial abundances. (In
the right-hand panel, the various theoretical and GC results have been plotted
as a function of [Fe/H] instead of $(V-I)_0^{\rm TRGB}$ to illustrate the near
constancy of $M_I^{\rm TRGB}$ over quite a wide range in metallicity, 
$\lta -2.5$ [Fe/H] $\lta -0.9$.)

Because of differences in the reference solar abundances, DSEP isochrones for
[$\alpha$/Fe] $= +0.4$, where $\alpha$ includes all of the $\alpha$ elements
(including oxygen), have almost the same absolute C$+$N$+$O abundance at a given
[Fe/H] value as V-R models for [O/Fe] $= +0.6$ and [$m$/Fe] $= +0.4$ for all
other $\alpha$ elements.  Accordingly, one would expect that they should predict
nearly the same stellar properties at the TRGB, and indeed they do, though
the dependence of $M_I^{\rm TRGB}$ on $(V-I)_0$ given by these models does not
follow the empirical relation (the solid orange curve in the left-hand panel)
as well as the predictions from V-R isochrones.  On the other hand, BaSTI
isochrones for [$\alpha$/Fe] $= +0.4$ assume a lower C$+$N$+$O abundance at a
fixed [Fe/H] than either the V-R or DSEP models; consequently, they would be
expected to lie somewhat below the black solid curve instead of $\approx
0.06$--0.07 mag above it in order to be consistent with the other results.

\citet{swc17} recently carried a study of the brightness of the TRGB, using
both the BaSTI and GARSTEC (\citealt{ws08}) stellar evolution codes, with a
focus on the input physics and the transformations of the models from the
H-R diagram to observational CMDs.  When updated nuclear reaction rates,
radiative and conductive opacities, neutrino cooling processes, etc. are
adopted, their computations predict $M_I^{\rm TRGB}$ magnitudes that are just
outside of the error bars of the JL17 calibration, on the high side.
V-R isochrones do not have this difficulty even though the same or very similar
basic physics has been incorporated into the Victoria code (this was checked).
The largest difference is probably associated with the equation of state (EOS),
but if the highly regarded FreeEOS\footnote{http://freeeos.sourceforge.net} is
adopted instead of the default EOS, the TRGB luminosity is increased by only
$\Delta\,\log L/L_\odot = 0.002$ (or 0.005 mag).  To further investigate this
issue, a prediction of the TRGB luminosity from the MESA code (\citealt{pbd11})
was obtained for a $0.8 \msol$ model with [Fe/H] $= -1.5$, [$\alpha$/Fe] $=
+0.4$, and $Y = 0.25$ from P.~Denisenkov (private communication).  This
differed from the value of $L^{\rm TRGB}$ that is obtained for the same case
using the Victoria code by only $\Delta\,\log L/L_\odot = 0.001$, despite small
differences in the adopted nuclear reactions and the treatment of diffusion.
MESA employs JINA nuclear reaction rates (\citealt{caf10}) and treats diffusion
using a code similar to that developed by \citet{tbl94}.

In their presentation of a new method to measure TRGB magnitudes, \citet{dbd20}
also examined some recent theoretical predictions of the TRGB luminosity;
specifically, those derived from PARSEC (\citealt{bmg12}, \citealt{mgb17}) and
MIST (\citealt{cdc16}) isochrones.  They found rather poor agreement between
these computations, with the PARSEC models predicting brighter absolute TRGB
magnitudes than the MIST isochrones by $\gta 0.15$ mag at a fixed colour; the
respective theoretical relations between $M_{F814}^{\rm TRGB}$ and the
$M_{F606W}-M_{F814W}$ colour bracketed the empirical relation by JL17.
The MIST models were generated using the MESA code, but as noted by Durbin et
al.~they assume scaled-solar abundances of the metals instead of a mixture with
enhanced abundances of the $\alpha$ elements.  It is clear from the left-hand
panel of Fig.~\ref{fig:f11} that a reduced O abundance results in fainter TRGB
absolute magnitudes, and that there is reasonable consistency between the MIST
and V-R predictions when the differences in the assumed chemical abundances are
taken into account.  In fact, it has already been mentioned in the preceeding
paragraph that the TRGB luminosities which are obtained from MESA and Victoria
stellar models are in very good agreement when both assume [$\alpha$/Fe]
$= +0.4$.

Small differences in the adopted stellar physics may not be primarily
responsible for the larger variations in the predicted values of $L^{\rm TRGB}$
mentioned above.  \citet{swc17} have already shown the importance of employing
very small timesteps in following the evolution along the RGB when the mass is
used as the independent variable.  It is also possible that the way in which the
stellar structure and the nucleosynthesis equations are solved could have an
impact on the predicted TRGB luminosities.  For instance, in the Victoria and
MESA codes, these equations are solved implicitly; i.e., the chemical abundances
and the structure variables (normally consisting of the radius, luminosity,
pressure, and temperature) are converged together so that the chemical profiles
and the stellar structure are always fully consistent with each other.

A different approach is used by the GARSTEC code (\citealt{ws08}), and possibly
some others, insofar as the abundance changes that occur in a given timestep
are calculated on the assumption of nuclear reaction rates based on the
abundances and thermodynamic properties of the gas given by the last converged
model, and then the stellar structure equations are solved without updating the
solution of the nucleosynthesis equations.\footnote{A similar procedure was used
in the computation of BaSTI stellar models until recently.  According to
S.~Cassisi (private communication), the current version of the BaSTI code, as
used by \citet{hpc18} and others since 2017, does iterate between the solution
of the stellar structure and nucleosynthesis equations during each timestep.}
To avoid significant errors, an explicit treatment of the chemical abundance
changes such as this requires the use of especially small timesteps.  Whether or
not this suggestion is part of the explanation for some of the discordant
results that are obtained from different codes is not known, but it would be
worthwhile to investigate this possibility. 

\begin{figure}
\begin{center}
\includegraphics[width=\columnwidth]{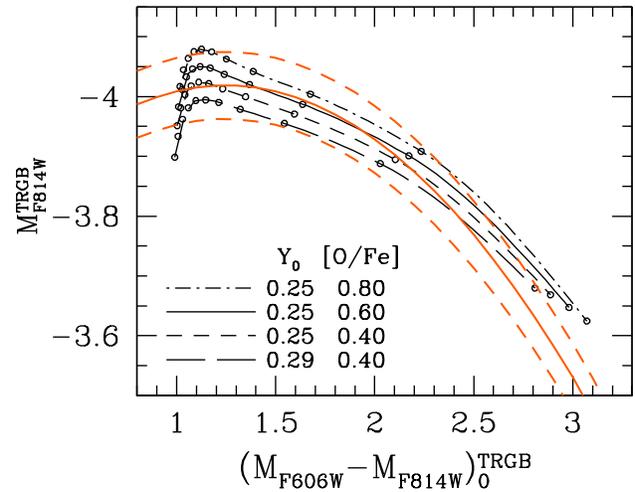}
\caption{As in the left-hand panel of Fig.~\ref{fig:f11}, except that the
theoretical results are compared with the JL17 empirical relation between 
$M_{F814W}^{\rm TRGB}$ and $(M_{F606W}-M_{F814W})_0$ colour.}
\label{fig:f13}
\end{center}
\end{figure}

In any event, it has been shown that the relation between $M_I^{\rm TRGB}$ and
$(V-I)_0$ colour given by V-R isochrones is in excellent agreement with 
empirical determinations from galactic and GC studies, and that similar 
consistency is found using MESA and DSEP models.  This success is not limited 
to $VI$ data.  As shown in Figure~\ref{fig:f13}, which considers ACS
$F606W,\,F814$ photometry, V-R isochrones also reproduce the observed relation
between $M_{F814W}^{\rm TRGB}$ and the $M_{F606W}-M_{F814W})_0$ colour index
to within its $1\,\sigma$ uncertainty over most of the colour range that has 
been plotted.  The observed colour dependence is not matched by the models quite
as well as in Fig.~\ref{fig:f11}, which is presumably a consequence of
systematic differences in the respective bolometric corrections as a function
of $\teff$.  A final point: given the considerable evidence in support of
Victoria ZAHB models, the He core mass and the envelope helium abundance at the
TRGB must be quite close to the values given by V-R isochrones since it is these
properties that largely determine where ZAHB models are located on the H-R
diagram.

\section{Summary}
\label{sec:sum}

Grids of stellar evolutionary tracks have been computed for different abundances
of C, N, and O (the mixtures listed in Table~\ref{tab:t1}) primarily to provide
a set of models that are relevant to the chemically distinct stellar populations
that reside in GCs.  Whereas Papers I and II considered stellar models for only
a few metallicities, the present grids were generated for sufficiently fine
spacings in [Fe/H] and $Y$ to permit the calculation of isochrones for any
metallicity and He abundance within the ranges $-2.5 \le$ [Fe/H] $\le -0.5$ and
$0.25 \le Y \le 0.29$, respectively.  These isochrones are limited to ages
$\gta 7$ Gyr because the evolutionary sequences were computed only for masses
from 0.12 to $1.0\,\msol$.  The same bolometric corrections (BCs) that were
presented in Paper I, which are based on fully consistent MARCS model
atmospheres and synthetic colours, are used to transform the isochrones to CMDs
that involve the magnitudes and colours which characterize many of the
broad-band photometric systems currently in use --- including, in particular,
Johnson-Cousins $UBVRI$ and most of the {\it HST} ACS and WFC3 filters.

Paper II has already demonstrated that V-R isochrones are reasonably successful
in fitting the UV/optical CMDs that are derived from the {\it HST} UV Legacy
Survey {\citealt{pmb15}, \citealt{nlp18}).  There are certainly some unexplained
features, such as the populations of lower RGB stars with very blue
$(M_{F336W}-M_{F438W})_0$ colours that seem to be common to all GCs, but the
observed colour spreads encompassed by most of the cluster stars are similar to
those predicted for CN-weak and CN-strong stars.  Moreover, the morphologies of
the observed CMDs are reproduced quite well by the models, including the
development of the hook feature in the vicinity of the TO in
$(M_{F336W}-M_{F438W})_0,\,M_{F606W}$ diagrams as the metallicity decreases
below [Fe/H] $\sim -2$.  While predicted $M_{F336W}$ magnitudes appear to be
too bright by about 0.03 mag, there are no obvious zero-point offsets between
the observed and synthetic magnitudes that are derived from redder passbands.

In this study, the comparisons with observations have been extended to the IR.
In general, V-R isochrones appear to provide good fits to the bluest stars
below the knee in, e.g., $(M_{F110W}-M_{F160W})_0,\,M_{F160W}$ diagrams,
indicating that they have [O/Fe] $\approx +0.6$, but not the reddest
populations.  Whether this is due to deficiencies in the BCs for mixtures with
very low O abundances or the stars have chemical properties that differ in some
significant way from those assumed in the present models is not known.  Errors
in the model $\teff$ scale seem unlikely given that the isochrones provide
excellent fits to observed $(M_{F606W}-M_{F814W})_0$ colours down to $M_{F606W}
\gta 8$ ($\sim 4$ magnitudes below the TO).  The IRFM temperatures that have
been derived by \citet{crm10} for TO and upper MS field halo stars with accurate
{\it Gaia} parallaxes, as well as their $(V-I_C)_0$ and $(V-K_S)_0$ colours, are
also well matched by the V-R isochrones.  The same computations predict TRGB
absolute magnitudes and their colour dependencies that agree very well with the
empirical results by \citet{jl17}.  Indeed, Victoria stellar models and the
isochrones derived from them appear to be particularly successful in satisfying
the available observational constraints. 

ZAHB models have played an important role in this investigation.  Grids of
such models for $-2.5 \le$ [Fe/H] $\le -0.5$ and at least two values of $Y$
and [O/Fe] at each metallicity will be provided in a forthcoming paper.

\section*{acknowledgements}
Sincere thanks go to Bengt Edvardsson for computing new model atmospheres for
[Fe/H] $= -2.0$ and $-1.0$ that were used as boundary conditions for models of
LMS stars, Pavel Denisenkov for computing a MESA track for comparison with one
generated by the Victoria code for the same parameters, Matteo Correnti and
Jason Kalirai for their IR photometry of 47 Tuc and NGC$\,$6397, Antonino Milone
for the IR observations of NGC$\,$6752 from the recent study by \citet{dmr22},
and Karsten Brogaard, Luca Casagrande, and Domenico Nardiello for helpful
information and/or comments.

\section*{Data Availability}
The evolutionary tracks for all of the metal abundance mixtures that have been
considered in this project, and the means to interpolate in these tracks to
produce isochrones for ages $\gta 7$ Gyr and for any metallicity and He abundance
within the respective ranges $-2.5 \le$ [Fe/H] $\le -0.5$ and $0.25 \le Y \le
0.29$, may be obtained from https://www.canfar.net/storage/list/VRmodels.  The
files relevant to this investigation are: (i) vriso.zip, which contains not
only the stellar models computed for this project, but also those presented by
\citet{vbf14} that allow for variations in [$\alpha$/Fe] at a given
metallicity, (ii) README\_vriso, which provides detailed instructions on how to
generate isochrones and to transform them to colour-magnitude diagrams using
user-friendly computer programs that are provided, and (iii) TRGB\_vriso.data,
which lists the properties of stellar models at the tip of the RGB (in small
tables similar to that shown in Table~\ref{tab:t2}), allowing for the effects
of varying $Y$ and [O/Fe] at metallicities ranging from $-2.5 \le$ [Fe/H] $\le
-0.5$ in steps of 0.2 dex.

\bsp   
\label{lastpage}

\end{document}